\documentclass[aps,pra,twocolumn,superscriptaddress,nofootinbib]{revtex4-1}

\usepackage{helvet}
\usepackage{eurosym}
\usepackage{color}
\usepackage[latin1]{inputenc}
\usepackage{amssymb}
\usepackage{amsmath}
\usepackage{graphicx} 
\usepackage{bm}
\usepackage{xcolor}
\usepackage{multirow}
\usepackage{dutchcal}

\usepackage{braket}

\usepackage{ulem}

\usepackage{natbib}

\usepackage{cancel}

\DeclareGraphicsExtensions{%
.png,%
.jpg,%
.pdf,.PDF,%
.mps,.jpeg,.jbig2,.jb2,.JPG,.JPEG,.JBIG2,.JB2}

\newcommand{\abs}[1]{\left| #1 \right|} 
\newcommand{\avg}[1]{\left< #1 \right>} 

\newcommand{\rb}[1]{\left( #1 \right)}

\newcommand{\trace}[1]{{\rm Tr}\left\{ #1 \right\}}

\DeclareMathAlphabet{\pazocal}{OMS}{zplm}{m}{n}

\newcommand{\ii}{{\rm i}}

\begin{document}
\title{Polaron-transformed dissipative Lipkin-Meshkov-Glick Model}

\author{Wassilij Kopylov}
\email{kopylov@itp.tu-berlin.de}
\author{Gernot Schaller}
\email{gernot.schaller@tu-berlin.de}
\affiliation{Institut f\"ur Theoretische Physik,
 Technische Universit\"at Berlin,
 D-10623 Berlin,
 Germany}
\date{\today}

\begin{abstract}
We investigate the Lipkin-Meshkov-Glick model coupled to a thermal bath.
Since the isolated model itself exhibits a quantum phase transition, we explore the critical signatures of the open system.
Starting from a system-reservoir interaction written in positive definite form, we find that 
the position of the critical point remains unchanged, in contrast to the popular mean-field prediction.
Technically, we employ the polaron transform  to be able to study 
the full crossover regime from the normal to the symmetry-broken phase,
which allows us to investigate the fate of quantum-critical points subject to dissipative environments.
The signatures of the phase transition are reflected in observables like magnetization, stationary mode occupation or waiting-time distributions.

\end{abstract}

\maketitle


\section{Introduction}

In closed systems, Quantum Phase Transitions (QPTs) are defined as non-analytic changes of the ground state energy when a control parameter other than temperature is varied across a critical point~\cite{Sachdev-QPT}. 
They are accompanied by non-analytic changes in observables or correlation functions~\cite{LMG-thermodynamical_limit-Mosseri,Hirsch-Dicke_TC_-quantum-and-semi-analysis-chaos,Dicke_Entanglement_and_QPT-Brandes} and form a fascinating research area on their own.

Nowadays, it is possible to study such QPTs in experimental setups with cold atoms~\cite{Baumann-Dicke_qpt,Baumann-symm-break-in-Dicke-QPT,brennecke2013real,LMG-Exp_Bifurcation_rabi_to_jesophson-Oberthaler,Ritsch-Domokos-Cold-atoms-in-opt-potential}, which provide high degree of control and allow to test theoretical predictions.
However, each experimental set-up is an open system, such that the impact of the reservoir on the QPT should not be neglected.
To the contrary, the presence of a reservoir can fundamentally change the nature of the QPT.
For example, in the famous Dicke phase transition, it is the presence of the reservoir that actually creates a QPT via the environmental coupling of a collective spin~\cite{Dicke-Dicke_Modell}. 

With the renewed interest in quantum thermodynamics, it has become a relevant question whether QPTs can be put to use e.g.\ as working fluids of quantum heat engines~\cite{fusco2016a,cakmak2016a,ma2017a,kloc2019a}.
This opens another broad research area of dissipative QPTs in non-equilibrium setups. 
Here, the non-equilibrium configuration can be implemented in different ways, e.g.\ by periodic driving~\cite{Dicke-nonequilibrium_qpt-bastidas,LMG-ac_driven_QPT-Georg,Bastidas-Critical_quasienergy_in_driven_systems}, by quenching~\cite{Dicke-Robust_quantum_correlation_with_linear_increased_coupling-Acevedo,LMG-Nonadiabatic_dynamics_of_ESQPT-kopylov,LMG-Criticality_revealed_and_quench_dynamics-Campbell}, by coupling to reservoirs~\cite{Scully-quantum_optics,LMG-collective_and_independent-decay-Lee,Kopylov_Counting-statistics-Dicke} or by a combination of these approaches~\cite{mostame2007a,mostame2010a}.
One has even considered feedback control of such quantum-critical systems~\cite{Dicke_Dynamical_phase_transition_open_Dicke-Klinder,Thermalization-Pseudo_NonMarkovian_dissipative_system-Chiocchetta,Feedback-mirror_propiertes_as_time_delay_fb-carmele,Kopylov-LMG_ESQPT_control,Pyragas-Uravaling_coherent_FB-Kabuss}. 

All these extensions should however be applied in combination with a reliable microscopic description of the system-reservoir interaction.
For example, in the usual derivation of Lindblad master equations one assumes that the system-reservoir interaction is weak compared to the splitting of the system energy levels~\cite{Scully-quantum_optics,Breuer-open_quantum_systems}.
In particular in the vicinity of a QPT -- where the energy gap above the ground state vanishes -- this condition cannot be maintained.
Therefore, while in particular the application of the secular approximation leads to a Lindblad-type master equation preserving the density matrix properties, it has the disadvantage that its range of validity is typically limited to non-critical points or to finite-size scaling investigations~\cite{vogl2012b,schaller2014a}.
In principle, the weak-coupling restriction can be overcome with different methods such as e.g.\ reaction-coordinate mappings~\cite{Schaller-QS_far_from_equilibrium,nazir2018a,Thermodynamics-Nonequilibr_react_coordinate-Strasberg}. 
These however come at the price of increasing the dimension of the system, which renders analytic treatments of already complex systems difficult.

In this paper, we are going to study at the example of the Lipkin-Meshkov-Glick (LMG) model how a QPT is turned dissipative by coupling the LMG system~\cite{LMG-validity_many_body_approx-Lipkin} to a large environment. 
To avoid the aforementioned problems, we use a polaron~\cite{mahan2013many,Polaron-inelastic_resonant_tunneling_electrons_barrier-Glazman,Polaron-electron_phonon_resonant_tunneling-wingreen,Polaron-coherent_collective_effects-_mesoscopic-Brandes,Polaron-electron_transistor_strong_coupling_counting-Schaller} method, which allows to address the strong coupling regime~\cite{Polaro-spin_boson_dynamics-Thorwart,Polaron-spin_boson_comparisson-Wilhelm,Polaron-dissipation_two_level_comparisson-brandes,Dicke_ultra-strong-coupling-limit-qpt_Bucher,Schaller-QS_far_from_equilibrium,Krause2015,Keeling_PRL-nonequilibrium_model_photon-cond,PhotCond-interplay_coh_disisp_dynamics-Milan} without increasing the number of degrees of freedom that need explicit treatment.
In particular, we show that for our model the position of the QPT is robust in presence of dissipation.
We emphasize that the absence of a reservoir-induced shift -- in contrast to mean-field-predictions~\cite{Bhaseen_dynamics_of_nonequilibrium_dicke_models,Dicke_open-critical_exponent_of_noise-Nagy,Kopylov_Counting-statistics-Dicke,Rabi_dissipative-QPT_Plenio,Dicke-dissipative_bistability_noise_nonthermal-Buchhold,Dicke-dynamical_symmetry_breaking-Li,Morrison-Dissipative_LMG-and_QPT} -- 
is connected with starting from a Hamiltonian with a lower spectral bound and holds without additional approximation.
Our work is structured as follows. 
In Sec.~\ref{sec:1} we introduce the dissipative LMG model, in Sec.~\ref{sec:2} we show how to diagonalize it globally using the Holstein-Primakoff transformation. 
There, we also derive a master equation in both, original and polaron, frames and show that the QPT cannot be modeled within the first and that the QPT position is not shifted within the latter approach.
Finally, we discuss the effects near the QPT  by investigating the excitations in the LMG system and the waiting time distribution of emitted bosons in Sec.~\ref{sec:3}. 


\section{Model}\label{sec:1}

\subsection{Starting Hamiltonian}

The isolated LMG model describes the collective interaction of $N$ two-level systems with an external field and among themselves. 
In terms of the collective spin operators 
\begin{align}
J_\nu = \frac{1}{2}\sum_{m=1}^N \sigma_\nu^{(m)}\,,\qquad \nu \in \{x,y,z\}
\end{align}
and $J_\pm = J_x \pm {\rm i} \cdot J_y$
with $\sigma_\nu^{(m)}$ denoting the Pauli matrix of the $m$th spin, 
the anisotropic LMG Hamiltonian reads~\cite{LMG-Critical_scaling_law_entaglement-Vidal}
\begin{equation}
\label{eq:LMG}
H_{\rm LMG}(h,\gamma_x) = -h J_z - \frac{\gamma_x}{N} J_x^2\,,
\end{equation}
where $h$ is the strength of a magnetic field in $z$ direction and $\gamma_x$ is the coupling strength between each pair of two-level systems. 
As such, it can be considered a quantum generalization of the Curie-Weiss model~\cite{kochmanski2013a}.
Throughout this paper, we consider only the subspace with the maximum angular momentum $j=\frac{N}{2}$, where the eigenvalues of the angular momentum
operator $J^2 = J_x^2 + J_y^2 + J_z^2$ are given by $j(j+1)$.
Studies of the LMG model are interesting not only due to its origin in the nuclear context~\cite{LMG-lipkin1965validity,LMG-validity_many_body_approx-Lipkin,LMG-validity_many_body_approx-Lipkin-3}, but also due to its experimental realization with cold atoms and high possibility of control~\cite{LMG-Exp_Bifurcation_rabi_to_jesophson-Oberthaler}. 
In particular the existence of a QPT at $\gamma_x^{\rm cr} = h$ with a non-analytic ground-state energy density has raised the interest in the community~\cite{LMG-phase_transition-Gilmore,LMG-large_N_scaling_behaviour-Heiss,LMG-networks_qpt-sorokin,LMG_Entanglement_dynamics_Vidal}: 
For $\gamma_x < \gamma_x^{\rm cr}$, the system has a unique ground state, which we denote as the {\it normal phase} further-on.
In contrast, for $\gamma_x > \gamma_x^{\rm cr}$ it exhibits a {\it symmetry-broken phase}~\cite{LMG-thermodynamical_limit-Mosseri,LMG-symmetry_breaking_dynamics_finite_size-Huang}, where e.g.\ the  eigenvalues become pairwise degenerate and the $J_z$-expectation exhibits a bifurcation~\cite{LMG-spectrum_thermodynamic_limit_and_finite_size-corr-Mosseri,LMG-Nonadiabatic_dynamics_of_ESQPT-kopylov}.
Strictly speaking, the QPT is found only in the thermodynamic limit (for $N \to \infty$), for finite sizes $N$ smoothing effects in the QPT signatures will appear~\cite{LMG-Finite_size_scalling_Dusuel,LMG-Finite_size_scaling-Vidal,LMG-Wiseman_control-Kopylov}.
   
Here, we want to investigate the LMG model embedded in an environment of bosonic oscillators $c_k$  with frequencies $\nu_k$.
The simplest nontrivial embedding preserves the conservation of the total angular momentum and allows for energy exchange between system and reservoir.
Here, we constrain ourselves for simplicity to the case of a $J^x$ coupling. 
Furthermore, to
ensure that the Hamiltonian has a lower spectral bound for all values of the system-reservoir coupling strength, we write the interaction in terms of a positive operator
\begin{align}
\label{eq:H_LMG_And_Bath_ini}
H_{\rm tot} &= H_{\rm LMG}(h,\gamma_x)\notag \\
	&\;+\sum_k \nu_k \rb{c_k^\dag + \frac{g_k}{\sqrt{N} \nu_k}J_x}\rb{c_k + \frac{g_k}{\sqrt{N} \nu_k}J_x}\,.
\end{align} 
Here, $g_k > 0$ represent emission/absorption amplitudes (a possible phase can be absorbed in the bosonic operators), and the factor $N^{-1/2}$ needs to be
included to obtain a meaningful thermodynamic limit $N \to \infty$, but can also be motivated from the scaling of the quantization volume $V \propto N$. 
Since the LMG Hamiltonian has a lower bound, the spectrum of this Hamiltonian $H_{\rm tot}$ is (for finite $N$) then bounded from below for all values of the
coupling strength $g_k$.
Upon expansion and sorting spin and bosonic operators, this form implicates an effective rescaling of the system Hamiltonian $H_{\rm LMG}(h,\tilde\gamma_x)$ with a renormalized spin-spin interaction 
\begin{equation}\label{EQ:interaction_rescaled}
\tilde{\gamma}_x = \gamma_x - \sum_k \frac{g_k^2}{\nu_k}\,,
\end{equation} 
which indeed leads to a shift of the critical point within a naive treatment.

\subsection{Local LMG diagonalization}

In the thermodynamic limit Eq.~\eqref{eq:LMG} can be diagonalized using the Holstein-Primakoff transform which maps collective spins to bosonic operators $b$~\cite{HP-Trafo_field_dependency_of_ferromagnet_Primakoff,Clive-Brandes_Chaos_and_qpt_Dicke,Kopylov_Counting-statistics-Dicke}
\begin{align}
\label{eq:HP_trafo}
J_+ &= \sqrt{N - b^\dag b} b\,, \qquad J_- = b^\dag \sqrt{N - b^\dag b}\,,\\  
J_z &= \frac{N}{2} - b^\dag b\,.\notag
\end{align}
However, to capture both phases of the LMG Hamiltonian, one has to account for the macroscopically populated ground state in the symmetry-broken phase. 
This can be included with the displacement $b = \sqrt{N}\alpha + a$ with complex $\alpha$ in Eq.~\eqref{eq:HP_trafo}, where $N\abs{\alpha}^2$ is the classical mean-field population of the mode~\cite{Clive-Brandes_Chaos_and_qpt_Dicke,Kopylov_Counting-statistics-Dicke,LMG-networks_qpt-sorokin}
and $a$ is another bosonic annihilation operator.
The next step is then to expand for either phase Eq.~\eqref{eq:LMG} with the inserted transformation~\eqref{eq:HP_trafo} in terms 
of $1/\sqrt{N}$ for $N\gg 1$ -- see App.~\ref{APP:tdlimit} -- which yields a decomposition of the Hamiltonian
\begin{align}
\label{eq:H_LMG_HP}
H_{\rm LMG}^{\rm HP}(h,\gamma_x) &= N \cdot H_0^{\rm HP} + \sqrt{N} H_1^{\rm HP} + H_2^{\rm HP}\\\notag
&\qquad + {\mathcal O}\left(\frac{1}{\sqrt{N}}\right)\,,
\end{align}
with individual terms depending on the phase
\begin{align}
H_0^{\rm HP} &= \begin{cases}
-\frac{h}{2} &: \gamma_x < \gamma_x^{\rm cr}\\
-\frac{h^2 + \gamma_x^2}{4\gamma_x} &: \gamma_x > \gamma_x^{\rm cr}
\end{cases}\,,\\
H_1^{\rm HP} &\stackrel{!}{=} \begin{cases}
0 &: \gamma_x < \gamma_x^{\rm cr}\\
0 &: \gamma_x > \gamma_x^{\rm cr}
\end{cases}\,,\notag\\
H_2^{\rm HP} &=\begin{cases}
(h - \frac{\gamma_x}{2})a^\dag a - \frac{\gamma_x}{4} (a^2 + {a^\dag}^2)-\frac{\gamma_x}{4} &: \gamma_x < \gamma_x^{\rm cr}\\
+\frac{5 \gamma_x - 3 h}{4} a^\dagger a
+\frac{3 \gamma_x - 5 h}{8} \left(a^2 + {a^\dag}^2\right) &: \gamma_x > \gamma_x^{\rm cr}\\
\qquad{+\frac{\gamma_x - 3 h}{8}}
\end{cases}\,.\notag
\end{align}
We demand in both phases that $H_1^{\rm HP}$ is always zero.
Technically, this enforces that only terms quadratic in the creation and annihilation operators occur in the Hamiltonian.
Physically, this enforces that we expand around the correct ground state, i.e.,
in the final basis, the ground state is the state with a vanishing quasiparticle number.
This requirement is
trivially fulfilled in the normal phase with $\alpha=0$ but requires a finite
real value of the mean-field $\alpha$ in the symmetry-broken phase~\cite{Clive-Brandes_Chaos_and_qpt_Dicke,Kopylov_Counting-statistics-Dicke,LMG-networks_qpt-sorokin}, altogether leading to a phase-dependent displacement
\begin{equation}\label{eq:mean_field}
\alpha(h,\gamma_x) = \sqrt{\frac{1}{2} \left(1 - \frac{h}{\gamma_x}\right)} \Theta(\gamma_x - h)\,,
\end{equation} 
which approximates $H_{\rm LMG}^{\rm HP}$ by a harmonic oscillator near its ground state.
Here we note that $-\alpha(h,\gamma_x)$ is also a solution.
The mean-field expectation value already allows to see the signature of the phase transition in the closed LMG model at $\gamma_x = h$, since $\alpha$ is only finite for $\gamma_x > h$ and is zero elsewhere. 

\begin{figure}
\includegraphics[width=0.95\columnwidth,clip=true]{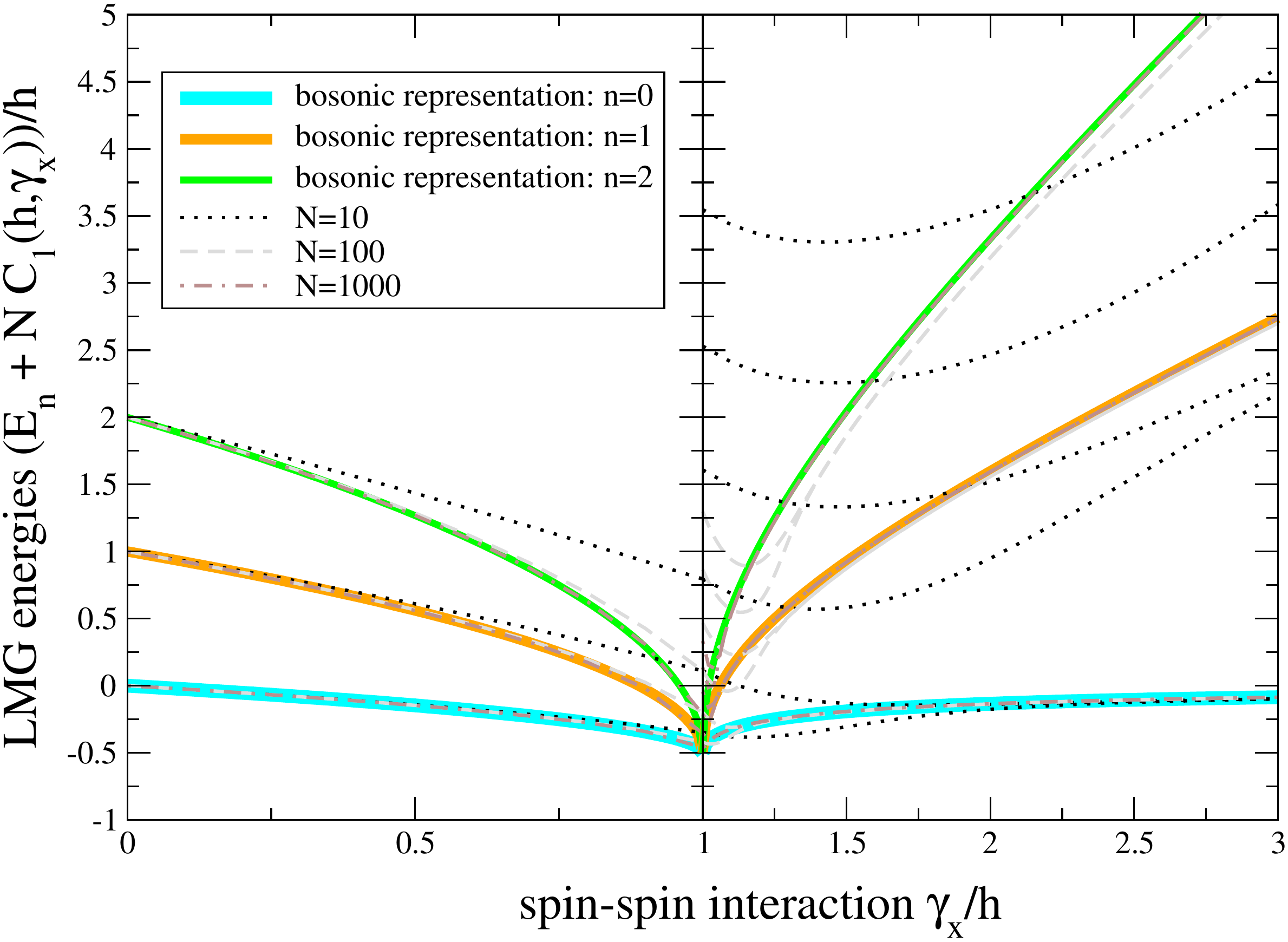}
\caption{\label{FIG:bogoliubov_lmg}
Lower part of the isolated LMG model spectrum for finite-size numerical diagonalization of Eq.~(\ref{eq:LMG}) (thin curves) and using the bosonic representation (bold curves) based on Eq.~(\ref{EQ:lmg_oscillator}) for the three lowest energies.
For large $N$, the spectra are nearly indistinguishable.
In the symmetry-broken phase (right), two numerical eigenvalues approach the same oscillator solution.
These correspond to the two different parity sectors, formally represented by two possible displacement solutions $\pm\alpha(h,\gamma_x)$ in Eq.~(\ref{eq:mean_field}).
}
\end{figure}

Since up to corrections that vanish in the thermodynamic limit, the Hamiltonian defined by Eq.~\eqref{eq:H_LMG_HP} is quadratic in $a$, it can in either phase be diagonalized by a rotation of the old operators $a=\cosh(\varphi) d+\sinh(\varphi) d^\dagger$ with $\varphi\in\mathbb{R}$ to new bosonic operators $d$. 
The system Hamiltonian $H_{\rm LMG}^{\rm HP}$ then transforms into a single harmonic oscillator, where the frequency $\omega$ and ground state energy are functions of $h$ and $\gamma_x$
\begin{align}\label{EQ:lmg_oscillator}
H_{\rm LMG}^{\rm HP}(h,\gamma_x) &= \omega(h,\gamma_x) d^\dag d + C_2(h,\gamma_x)\\\notag
&\qquad- N \cdot C_1(h,\gamma_x) 
+ {\mathcal O}\left(\frac{1}{\sqrt{N}}\right)\,.\notag
\end{align}
The actual values of the excitation energies $\omega(h,\gamma_x)$ and the constants $C_i(h,\gamma_x)$ are summarized in table~\ref{tab:0}. 
\begin{table}
$
\begin{array}{|l||l|l|}
\hline
 &	\text{normal:}\; \gamma_x<h 	&  \text{symmetry-broken:}\; \gamma_x>h \\ \hline\hline
b &	\multicolumn{2}{|c|}{\sqrt{N} \alpha(h,\gamma_x) + \cosh(\varphi(h,\gamma_x)) d + \sinh(\varphi(h,\gamma_x)) d^\dag} \\ \hline
\varphi(h,\gamma_x) & \frac{1}{4} \ln\left(\frac{h}{h-\gamma_x}\right)	&
\frac{1}{4} \ln\left(\frac{\gamma_x+h}{4(\gamma_x - h)}\right)\\ \hline
\alpha(h,\gamma_x) & 0 & \sqrt{\frac{1}{2}\left(1 - \frac{h}{\gamma_x}\right)} \\\hline
\omega(h,\gamma_x) & \sqrt{h(h-\gamma_x)} & 
\sqrt{\gamma_x^2 - h^2} \\ \hline
C_1(h,\gamma_x)	& \frac{h}{2}	& \frac{h^2 + \gamma_x^2}{4 \gamma_x} \\ \hline
C_2(h,\gamma_x) & \frac{1}{2} \left(\sqrt{h(h - \gamma_x)} - h\right) &
\frac{1}{2} \left(\sqrt{\gamma_x^2-h^2}-\gamma_x\right) \\ \hline
\end{array}	
$
\caption{Parameters of the diagonalization procedure of the LMG model $H_{\rm LMG}(h,\gamma_x)$ for the normal phase ($\gamma_x<h$, second column) and for the symmetry-broken phase ($\gamma_x>h$, last column). 
In both phases, the $d$ operators correspond to fluctuations around the mean-field value $\alpha$, which is zero only in the normal phase.
}
\label{tab:0}
\end{table}
Fig.~\ref{FIG:bogoliubov_lmg} confirms that the thus obtained spectra from the bosonic representation agree well with finite-size numerical diagonalization when $N$ is large enough.

First, one observes for consistency that the trivial spectra deeply in the normal phase ($\gamma_x \approx 0$) or deeply in the symmetry-broken phase ($h \approx 0$) are reproduced.
In addition, we see that at the QPT $\gamma_x = \gamma_x^{\rm cr}=h$, the excitation frequency $\omega$ vanishes as expected, which is also reflected
e.g.\ in the dashed curve in Fig.~\ref{fig:fluctuations}(a). 
For consistency, we also mention that all oscillator energies $E_n$ are continuous at the critical point $\gamma=h$.
Furthermore, the second derivative with respect to $\gamma_x$ of the continuum ground state energy per spin $\lim_{N\to\infty} E_0/N$ is discontinuous at the critical point, classifying the phase transition as second order.
Finally, we note that this treatment does not capture the excited state quantum phase transitions present in the LMG model as we are only concerned with the lower part of the spectrum.

\section{Master Equation} 
\label{sec:2}

We first perform the derivation of the conventional Born-Markov-secular (BMS) master equation in the usual way, starting directly with Eq.~\eqref{eq:H_LMG_And_Bath_ini} \cite{Kopylov_Counting-statistics-Dicke,LMG-collective_and_independent-decay-Lee,LMG_thermalization_kastner}.
Afterwards, we show that a polaron transform also allows to treat regions near the critical point.

\subsection{Conventional BMS master equation}

The conventional BMS master equation is derived in the energy eigenbasis of the system, i.e., the LMG model with renormalized spin-spin interaction $\tilde\gamma_x$, in order to facilitate the secular approximation. 
In this eigenbasis the master equation has a particularly simple form. 

Applying the very same transformations (that diagonalize the closed LMG model) to its open version~\eqref{eq:H_LMG_And_Bath_ini}, we arrive at
the generic form
\begin{align}
\label{eq:H_LMG_And_Bath_HP}
H_{\rm tot}^{\rm HP} &= H_{\rm LMG}^{\rm HP}(h,\tilde{\gamma}_x) + \sum \nu_k c_k^\dag c_k\notag \\
&\qquad+ \left[A(h,\tilde\gamma_x) (d + d^\dag) + \sqrt{N} Q(h,\tilde\gamma_x)\right] \times\notag\\
&\qquad\qquad \times \sum_k g_k (c_k + c_k^\dagger)\,,
\end{align}
where we note that the LMG Hamiltonian is now evaluated at the shifted interaction~(\ref{EQ:interaction_rescaled}).
The phase-dependent numbers $A$ and $Q$ are defined in Table~\ref{tab:1}.
\begin{table}
$
\begin{array}{|l||l|l|}
\hline
 &	\text{normal:}\; \tilde\gamma_x<h 	&  \text{symmetry-broken:}\; \tilde\gamma_x>h \\ \hline\hline
C_3(h,\tilde\gamma_x) & 1 & \frac{\sqrt{2} h}{\sqrt{\tilde\gamma_x(\tilde\gamma_x + h)}} \\ \hline
A(h,\tilde\gamma_x) & \multicolumn{2}{|c|}{\frac{C_3(h,\tilde\gamma_x)}{2} \exp[\varphi(h,\tilde\gamma_x)]} \\ \hline
Q(h,\tilde\gamma_x) & \multicolumn{2}{|c|}{\alpha(h,\tilde\gamma_x) \sqrt{1 - \alpha^2(h,\tilde\gamma_x)}} \\ \hline
\end{array}	
$
\caption{Additional parameters of the diagonalization procedure for the derivation of the master equation 
in the original frame for the normal phase ($\tilde\gamma_x < h$, second column) and for the symmetry-broken phase ($\tilde\gamma_x > h$, last column). 
Note that as compared to the closed model in Tab.~\ref{tab:0}, functions are evaluated at the shifted interaction~(\ref{EQ:interaction_rescaled}).
}
\label{tab:1}
\end{table}
In particular, in the normal phase we have $Q=0$, and we recover the standard problem of a harmonic oscillator weakly coupled to a thermal reservoir.
In the symmetry-broken phase we have $Q \neq 0$, such that the shift term in the interaction Hamiltonian formally diverges as $N\to\infty$, and a naive perturbative treatment does not apply.
Some thought however shows, that this term can be transformed away by applying yet another displacement for both system and reservoir modes
$d \to d + \sigma$ and $c_k \to c_k + \sigma_k$ with $\sigma,\sigma_k\in\mathbb{C}$ chosen such that all terms linear in creation and annihilation operators vanish in the total Hamiltonian.
This procedure does not change the energies of neither system nor bath operators, such that eventually, the master equation in the symmetry-broken phase is formally equivalent to the one in the normal phase, and the interaction proportional to $Q$ is not problematic.

Still, when one approaches the critical point from either side, the system spacing $\omega$ closes in the thermodynamic limit, which makes the interaction Hamiltonian at some point equivalent or even stronger than the system Hamiltonian.
Even worse, one can see that simultaneously, the factor $A \sim e^{+\varphi}$ in the interaction Hamiltonian diverges at the critical point,
such that a perturbative treatment is not applicable there.
Therefore, one should consider the results of the naive master equation in the thermodynamic limit $N\to\infty$ with caution.
The absence of a microscopically derived master equation near the critical point is a major obstacle in understanding the fate of quantum criticality in
open systems.

Ignoring these problems, one obtains a master equation having the standard form for a harmonic oscillator coupled to a thermal reservoir
\begin{align}\label{EQ:density_matrix}
\dot\rho(t) &= - \ii \left[H_{\rm LMG}^{\rm HP}(h,\tilde{\gamma}_x),\rho \right]
+ F_e \mathcal{D}(d)\rho + F_a \mathcal{D}(d^\dag)\rho\,,\notag\\
F_e &=  A^2(h,\tilde\gamma_x) \Gamma(\omega(h,\tilde\gamma_x)) [1 + n_B(\omega(h,\tilde\gamma_x)]\,,\notag\\
F_a &=  A^2(h,\tilde\gamma_x) \Gamma(\omega(h,\tilde\gamma_x)) n_B(\omega(h,\tilde\gamma_x))\,.
\end{align}
Here, we have used the superoperator notation 
$\mathcal{D}(O)\rho \hat{=} O \rho O^\dag - \frac{1}{2} \rho O^\dag O -\frac{1}{2} O^\dag O \rho$ for any operator $O$ and 
\begin{equation}
\label{eq:spectral_density_general}
\Gamma(\omega) = 2 \pi \sum_k g_k^2 \delta(\omega - \nu_k)
\end{equation} 
is the original spectral density of the reservoir, and 
$n_B(\omega)=[e^{\beta \omega}-1]^{-1}$ is the Bose distribution 
with inverse reservoir temperature $\beta$.
These functions are evaluated at the system transition frequency $\omega(h,\tilde\gamma_x)$.
The master equation has the spontaneous and stimulated emission terms in $F_e$ and the absorption term in $F_a$, and due to the balanced Bose-Einstein function these will at steady state just thermalize the system at the reservoir temperature, as is generically found for such BMS master equations. 
Note that $H_{\rm LMG}^{\rm HP}$ from Eq.~\eqref{EQ:density_matrix} is evaluated at the rescaled coupling
$\tilde\gamma_x$.
Therefore, the position of the QPT is at $\tilde\gamma_x^{\rm cr} = h$ and shifted to higher $\gamma_x$ couplings, see~\eqref{EQ:interaction_rescaled}.
Similar shifts of the QPT position in dissipative quantum optical models are known e.g.\ from mean-field treatments~\cite{Bhaseen_dynamics_of_nonequilibrium_dicke_models,Dicke-realization_Dicke_in_cavity_system-Dimer}. 
However, here we emphasize that we observe them as a direct consequence of ignoring  the divergence of interaction around the phase transition in combination with positive-definite form of the initial total Hamiltonian Eq.~\eqref{eq:H_LMG_And_Bath_ini}.

\subsection{Polaron master equation}

In this section, we apply a unitary polaron transform to the complete model, which has for other (non-critical) models been used to investigate the full regime of system-reservoir coupling strengths~\cite{Polaron-spin_boson-Wang,Polaron_collective_system_with_interaction-Wang}.
We will see that for a critical model, it can -- while still bounded in the total coupling strength -- be used to 
explore the systems behaviour at the QPT position.

\subsubsection{Polaron transform}

We choose the following polaron transform $U_p$ 
\begin{equation}
\label{eq:Polaron_trafo}
U_p = e^{-J_x \hat{B}}\,,\qquad \hat{B} = \frac{1}{\sqrt{N}} \sum_{k} \frac{g_k}{\nu_k} \left(c_k^\dag - c_k\right)\,.
\end{equation}
The total Hamiltonian~\eqref{eq:H_LMG_And_Bath_ini} in the polaron frame then becomes
\begin{align}
\label{eq:H_LMG_And_Bath_Polaron}
\bar{H}_{\rm tot} &= U_p^\dag H_{\rm tot} U_p\\
&= - h  D \cdot J_z - \frac{\gamma_x}{N} J_x^2 
+ \sum_k \nu_k c_k^\dag c_k\notag\\
&\qquad- h \cdot \left[J_z \cdot \left(\cosh(\hat{B}) - D\right) - \ii J_y \sinh(\hat{B})\right]\,. \notag
\end{align}
Here, $\gamma_x$ is the original interaction of the local LMG model, 
and the renormalization of the external field $D$ is defined via 
\begin{align}
\label{eq:bath_shift_polaron}
D &= \avg{\cosh (\hat{B})} = \trace{\cosh(\hat{B}) \frac{e^{-\beta \sum_k \nu_k c_k^\dagger c_k}}{\trace{e^{-\beta \sum_k \nu_k c_k^\dagger c_k}}}}\notag\\
&= \exp\left[-\frac{1}{N}\sum_k \left(\frac{g_k}{\nu_k}\right)^2 \left(n_k + \frac{1}{2}\right) \right] > 0\,,\notag\\
n_k &= \frac{1}{e^{\beta \nu_k} - 1}\,.
\end{align}
It has been introduced to enforce that the expectation value of the system-bath coupling vanishes for the thermal reservoir state.
More details on the derivation of Eq.~\eqref{eq:H_LMG_And_Bath_Polaron} are presented in App~\ref{app:polaron_derivation}.

The operator $\hat{B}\propto\frac{1}{\sqrt{N}}$ decays in the thermodynamic limit, such that for these studies, only the first few terms in the expansions of the $\sinh(\hat{B})$ and $\cosh(\hat{B})$ terms need to be considered. 

Accordingly, the position of the QPT in the polaron frame is now found at the QPT of the closed model 
\begin{equation}
\label{eq:qpt_position_polaron}
\gamma_x^{\rm cr} = h D 
\stackrel{N\to\infty}{\to} h\,.
\end{equation} 
Here, we have with $D\to 1$ implicitly assumed that the thermodynamic limit is performed in the system first.
If a spectral density is chosen that vanishes faster than quadratically for small frequencies, the above replacement holds unconditionally (see below).

We emphasize again we observe the absence of a QPT shift as a result of a proper system-reservoir interaction with a lower spectral bound. 
Without such an initial Hamiltonian, the reservoir back-action would shift the dissipative QPT~\cite{Bhaseen_dynamics_of_nonequilibrium_dicke_models,Dicke-realization_Dicke_in_cavity_system-Dimer}.

For the study of strong coupling regimes, polaron transforms have also been applied e.g.\ to single spin 
systems~\cite{Polaron-spin_boson-Wang} and collective non-critical spin systems~\cite{Polaron_collective_system_with_interaction-Wang}. 
Treatments without a polaron transformation should be possible in our case too, by rewriting Eq.~\eqref{eq:H_LMG_And_Bath_ini} in terms of reaction coordinates~\cite{Reaction_Coordinate-Effect_friction_electron_transfer_biomol-Garg,Thermodynamics-Nonequilibr_react_coordinate-Strasberg,nazir2018a}, leading to an open Dicke-type model.
  
In the thermodynamic limit, we can use that the spin operators $J_\nu$ scale at worst linearly in $N$ to expand the interaction and $D$, yielding
\begin{align}
\bar{H}_{\rm tot} &\approx - h  \left[1-\frac{1}{N} \delta\right] \cdot J_z - \frac{\gamma_x}{N} J_x^2 
+ \sum_k \nu_k c_k^\dag c_k\notag\\
&\qquad- h \cdot \left[\frac{J_z}{N} \left(\frac{1}{2} \bar{B}^2 + \delta\right) - \ii \frac{J_y}{\sqrt{N}} \bar{B}\right]\notag\\
&= -h J_z - \frac{\gamma_x}{N} J_x^2 + \sum_k \nu_k c_k^\dag c_k\notag\\
&\qquad- h \cdot \left[\frac{J_z}{N} \frac{1}{2} \bar{B}^2 - \ii \frac{J_y}{\sqrt{N}} \bar{B}\right]\,,
\end{align}
where $\bar{B} = \sqrt{N} \hat{B}$ and 
$D \equiv e^{-\frac{\delta}{N}}$ has been used.
As in the thermodynamic limit, $J_z/N$ just yields a constant, the first term in the last row can be seen as an all-to-all interaction between the environmental oscillators, which only depends in a bounded fashion on the LMG parameters $h$ and $\gamma_x$.
Since it is quadratic, it can be formally transformed away by a suitable global Bogoliubov transform
$c_k = \sum_q (u_{kq} b_q + v_{kq} b_q^\dagger)$
of all reservoir oscillators, which results in
\begin{align}
\bar{H}_{\rm tot} &\approx -h J_z - \frac{\gamma_x}{N} J_x^2 + \sum_k \tilde\nu_k b_k^\dag b_k\notag\\
&\qquad+ h \frac{\ii J_y}{\sqrt{N}} \sum_k \left(h_k b_k - h_k^* b_k^\dagger\right)\,,
\end{align}
and where $h_k \in \mathbb{C}$ are the transformed reservoir couplings and the $\tilde\nu_k$ the transformed reservoir energies. 
In case of weak coupling to the reservoir which is assumed here however, we will simply neglect the $\bar{B}^2$-term since it is then much smaller than the linear $\bar{B}$ term.  

\subsubsection{System Hamiltonian diagonalization}

To proceed, we first consider the normal phase $\gamma_x < h$.
We first apply the Holstein-Primakoff transformation to the total Hamiltonian, compare appendix~\ref{APP:tdlimit}.
Since in the normal phase the vanishing displacement implies $a=b$, this yields
\begin{align}
\label{eq:H_tot_Polaron_HP_Normal}
\bar{H}_{\rm tot, N}^{\rm (HP)} &= - \frac{h}{2} N  + \left(h-\frac{\gamma_x}{2}\right) a^\dag a -\frac{\gamma_x}{4}  ({a^\dag}^2 + a^2+1) 
\notag\\
&\quad + \sum_k \tilde{\nu}_k b_k^\dag b_k + \frac{h}{2}(a - a^\dag) \sum_k \left(h_k b_k - h_k^* b_k^\dagger\right)\,.
\end{align}
Here, the main difference is that the system-reservoir interaction now couples to the momentum of the LMG oscillator mode and not the position.
Applying yet another Bogoliubov transform $a = \cosh(\varphi(h,\gamma_x)) d + \sinh(\varphi(h,\gamma_x)) d^\dagger$ with the same parameters as in table~\ref{tab:0} eventually yields a Hamiltonian of a single diagonalized oscillator coupled via its momentum to a reservoir.

Analogously, the symmetry-broken phase $\gamma_x > h$ is treated with a finite displacement as outlined in App.~\ref{APP:tdlimit}.
The requirement, that in the system Hamiltonian all terms proportional to $\sqrt{N}$ should vanish, yields 
the same known displacement~(\ref{eq:mean_field}).
One arrives at a Hamiltonian of the form
\begin{align}
\label{eq:H_tot_Polaron_HP_SR}
\bar{H}_{\rm tot, S}^{\rm (HP)} &= -\frac{h^2+\gamma_x^2}{4\gamma_x} N + 
\frac{5 \gamma_x - 3 h}{4} a^\dagger a\\
&\qquad+\frac{3 \gamma_x - 5 h}{8} \left(a^2 + {a^\dag}^2\right)
+\frac{\gamma_x - 3 h}{8}
+ \sum_k \tilde\nu_k b_k^\dag b_k\notag\\
&\qquad+ \frac{h}{2} \sqrt{1-\abs{\alpha(h,\gamma_x)}^2} (a-a^\dagger) \sum_k (h_k b_k - h_k^* b_k^\dagger)\,.
\notag
\end{align}
Using a Bogoliubov transformation to new bosonic operators $d$ the system part in the above equation can be diagonalized again.

Thus, in both phases the Hamiltonian acquires the generic form
\begin{align}
\bar{H}_{\rm tot}^{\rm (HP)} &= \omega(h,\gamma_x) d^\dagger d - N C_1(h,\gamma_x) + C_2(h,\gamma_x)\notag\\
&\qquad+ \bar{A}(h,\gamma_x) (d-d^\dagger) \sum_k \left(h_k b_k - h_k^* b_k^\dagger\right)
\notag\\
&\qquad+ \sum_k \tilde{\nu}_k b_k^\dag b_k\,,
\end{align}
where the system-reservoir coupling modification $\bar{A}(h,\gamma_x)$ is found in Tab.~\ref{tab:2}.
\begin{table}
$
\begin{array}{|l||l|l|}
\hline
 &	\text{normal:}\; \gamma_x<h 	&  \text{symmetry-broken:}\; \gamma_x >h \\ \hline\hline
\bar{C}_3(h,\gamma_x) & h & h \sqrt{\frac{1}{2}\left(1 + \frac{h}{\gamma_x}\right)} \\ \hline
\bar{A}(h,\gamma_x) & \multicolumn{2}{|c|}{\frac{\bar{C}_3(h,\gamma_x)}{2} \exp[-\varphi(h,\gamma_x)]} \\ \hline
\end{array}	
$
\caption{Additional parameters of the diagonalization procedure of $H_{\rm LMG}$ in the polaron frame for the normal phase ($\gamma_x < h$, second column) and symmetry broken phase ($\gamma_x > h$, last column). 
Note that $\varphi(h,\gamma_x)$ -- see Tab.~\ref{tab:0} -- is evaluated at the original spin-spin coupling $\gamma_x$.}
\label{tab:2}
\end{table}	

To this form, we can directly apply the derivation of the standard quantum-optical master equation.

\subsubsection{Master Equation}

In the polaron-transformed interaction Hamiltonian, we do now observe the factor $\bar{A}(h,\gamma_x)$, which depends on $h$ and $\gamma_x$, see tables~\ref{tab:2} and~\ref{tab:0}.
This factor is suppressed as one approaches the shifted critical point, it vanishes there identically.
Near the shifted QPT, its square  $\bar{A}^2(h,\gamma_x)$ shows the same scaling behaviour as the system gap $\omega(h,\gamma_x)$, such that in the polaron frame, the system-reservoir interaction strength is adaptively scaled down with the system Hamiltonian, and a naive master equation approach can be applied in this frame. 

From either the normal phase or the symmetry-broken phase we arrive at the following generic form of the system density matrix master equation
\begin{align}
\label{eq:density_matrix_polaron}
\dot\rho(t) &= - \ii \left[H_{\rm LMG}^{\rm HP}(h,\gamma_x),\rho \right]
+ \bar{F}_e \mathcal{D}(d)\rho + \bar{F}_a \mathcal{D}(d^\dag)\rho\,,\notag\\
\bar{F}_e &=  \bar{A}^2(h,\gamma_x) \bar\Gamma(\omega(h,\gamma_x)) [1 + n_B(\omega(h,\gamma_x))]\,,\notag\\ 
\bar{F}_a &=  \bar{A}^2(h,\gamma_x) \bar\Gamma(\omega(h,\gamma_x)) n_B(\omega(h,\gamma_x))\,. 
\end{align}
Here, $\bar\Gamma(\omega) = 2 \pi \sum_k \abs{h_k}^2 \delta(\omega - \tilde\nu_k)$ denotes the transformed spectral density, which is related to the original spectral density via the Bogoliubov transform that expresses the $c_k$ operators in terms of the $b_k$ operators, and $n_B(\omega)$ again denotes the 
Bose distribution.
The mapping from the reservoir modes $c_k$ to the new reservoir modes $b_k$ has been represented in an implicit form, but in general it will be a general multi-mode Bogoliubov transformation~\cite{Hamilt_diagonalization_quadratik_Tsallis1978,Hamilt_diagonalization_quadratik_Tikochinsky}
with a sophisticated solution.

However, if $h g_k/\nu_k$ is small in comparison to the reservoir frequencies $\nu_k$, the Bogoliubov transform will hardly change the reservoir oscillators and thereby be close to the identity.
Then, one will approximately recover 
$\bar\Gamma(\omega) \approx \Gamma(\omega)$.
Even if this assumption is not fulfilled, we note from the general form of the master equation that the steady state will just be the thermalized system -- with renormalized parameters depending on $\Gamma(\omega)$, $h$, and $\gamma_x$.
Therefore, it will not depend on the structure of $\bar\Gamma(\omega)$ -- although transient observables may depend on this transformed spectral density as well.
In our results, we will therefore concentrate on a particular form of $\Gamma(\omega)$ only and neglect the implications for $\bar\Gamma(\omega)$.

\section{Results}
\label{sec:3}

To apply the polaron transform method, we require that all involved limits converge.
All reasonable choices for a spectral density~(\ref{eq:spectral_density_general}) will lead to convergence of the renormalized spin-spin interaction~(\ref{EQ:interaction_rescaled}).
However, convergence of the external field renormalization~(\ref{eq:bath_shift_polaron}) may require subtle discussions on the order of the thermodynamic limits in system ($N\to\infty$)
and reservoir ($\sum_k g_k^2 [\ldots] \to \frac{1}{2\pi}\int \Gamma(\omega)[\ldots]d\omega$), respectively.
These discussions can be avoided if the spectral density grows faster than quadratically for small energies, e.g.
\begin{equation}
\label{eq:specral_density}
\Gamma(\omega)  =\eta \frac{\omega^3}{\omega_c^2} \cdot \exp(-\omega/\omega_c)\,,
\end{equation}
where $\omega_c$ is a cutoff frequency and $\eta$ is a dimensionless coupling strength. 
With this choice, the renormalized all-to-all interaction~(\ref{EQ:interaction_rescaled}) becomes
\begin{equation}
\tilde{\gamma}_x = \gamma_x - \frac{\eta \cdot \omega_c}{\pi}\,,
\end{equation}   
such that the QPT position Eq.~\eqref{EQ:interaction_rescaled} is shifted to 
$\gamma_x^{\rm cr} \to h + \frac{\eta \cdot \omega_c}{\pi}$.

We emphasize again that -- independent of the spectral density -- both derived master equations Eq.~\eqref{EQ:density_matrix} and~\eqref{eq:density_matrix_polaron} let the system evolve towards the respective thermal state
\begin{equation}
\label{eq:density_matrix_steady_state}
\rho = \frac{\exp(-\beta H_{\rm LMG}^{\rm HP}(h,\tilde\gamma_x))}{Z}\,,\;\;
\bar\rho = \frac{\exp(-\beta H_{\rm LMG}^{\rm HP}(h,\gamma_x))}{\bar{Z}}\,,
\end{equation}
in the original and polaron frame, respectively,
where $\beta$ is the inverse temperature of the bath and $Z/\bar{Z}$ are the respective normalization constants. 

The difference between the treatments is therefore that within the BMS treatment~\eqref{EQ:density_matrix} the rates may diverge and that the system parameters are renormalized.
The divergence of rates within the BMS treatment would also occur for a standard initial Hamiltonian.
To illustrate this main result, we discuss a number of conclusions can be derived from it below.

\subsection{Magnetization}

\begin{figure}
\includegraphics[width=0.95\columnwidth]{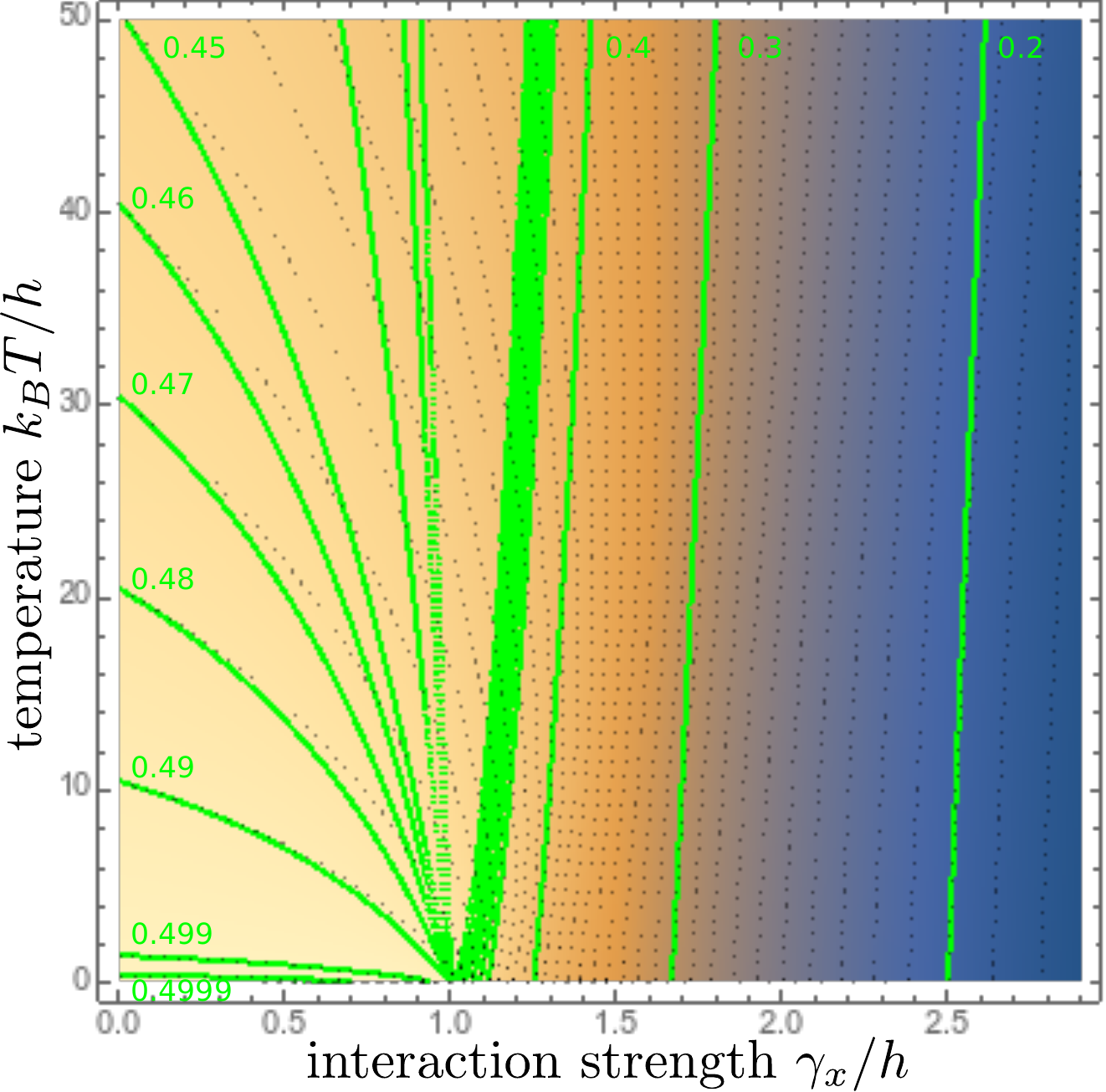}
\caption{\label{FIG:magnetization}
Contour plot of the magnetization density $\avg{J^z}/N$ versus spin-spin interaction $\gamma_x$ and temperature $k_B T$ for $N=1000$.
At the critical point $\gamma_x = h$, the magnetization density at low temperatures (bottom) suddenly starts to drop from a constant value in the normal phase (left) to a decaying curve in the symmetry-broken phase (right) as predicted by~(\ref{EQ:mag_gs}).
At higher temperatures, the transition is smoother and the predictions from the bosonic representation (solid green contours, based on Eq.~(\ref{EQ:mag_ss}))
and the finite-size numerical calculation of the partition function (dashed contours, based on the Gibbs state with Eq.~(\ref{eq:LMG})) disagree for $\gamma_x \approx h$.
For the finite-size calculation, weak coupling has been assumed $k_B T \ll N \omega_c/\eta $, such that $U_p^\dag J_z U_p \approx J_z$ instead of   \eqref{eq:jz_polaron_non_polaron_connection}.
}
\end{figure}

In general, the role of temperature in connection with the thermal phase transition in models like LMG or Dicke has been widely studied using partition sums or by using naive BMS master equations
\cite{LMG-thermal_phase_partition_sum_Tzeng,hayn2017thermodynamics,wilms2012finite,dalla2016dicke}.
Since in our case the stationary system state is just the thermalized one, standard methods (compare Appendix~\ref{APP:magnetization}) just analyzing the canonical Gibbs state of the isolated LMG model can be used to obtain stationary expectation values such as e.g. the magnetization. 
For the  polaron approach we obtain
\begin{align}\label{EQ:mag_ss}
\avg{J^z} = -\frac{\partial E_0(h,\gamma_x)}{\partial h} - \frac{1}{e^{\beta \omega(h,\gamma_x)}-1} \frac{\partial\omega(h,\gamma_x)}{\partial h}\,,
\end{align}
where $E_0(h,\gamma_x)=C_2(h,\gamma_x)-N C_1(h,\gamma_x)$ is the ground state energy and $\omega(h,\gamma_x)$ the energy splitting, compare Tab.~\ref{tab:0}.
The quantum-critical nature is demonstrated by the first (ground state) contribution, where the nonanalytic dependence of the ground state energy on the external field strength will map to the magnetization.
The second contribution is temperature-dependent.
In particular, in the thermodynamic limit $N\to\infty$, only a part of the ground state contribution remains and we obtain
\begin{align}\label{EQ:mag_gs}
\lim_{N\to\infty}\frac{\avg{J^z}}{N} \to \frac{1}{2}\left\{\begin{array}{ccc}
1 &:& h > \gamma_x\\
\frac{h}{\gamma_x} &:& \gamma_x>h
\end{array}\right.\,.
\end{align}
For finite system sizes however, finite temperature corrections exist.
In Fig.~\ref{FIG:magnetization}, we show a contour plot of the magnetization density $\avg{J^z}/N$ from the exact numerical calculation of the partition function (dashed contours) and compare with the results from the bosonic representation (solid green contours).
We see in the contour lines of the magnetization convincing agreement between the curves of the bosonic representation (solid green) and the finite-size calculation (dashed black) only for very low temperatures or away from the critical point.
The disagreement for $\gamma_x \approx h$ and $T>0$ can be attributed to the fact that the bosonization for finite sizes only captures the lowest energy eigenstates well, whereas in this region also the higher eigenstates become occupied.
However, it is clearly visible that in the low temperature regime, the magnetization density will drop suddenly when $\gamma_x \ge h$, such that the QPT can be detected at correspondingly low temperatures.
At high temperatures, the magnetization density falls of smoothly with increasing spin-spin interaction.

\subsection{Mode Occupation}

\begin{figure}
\includegraphics[width=0.51 \columnwidth]{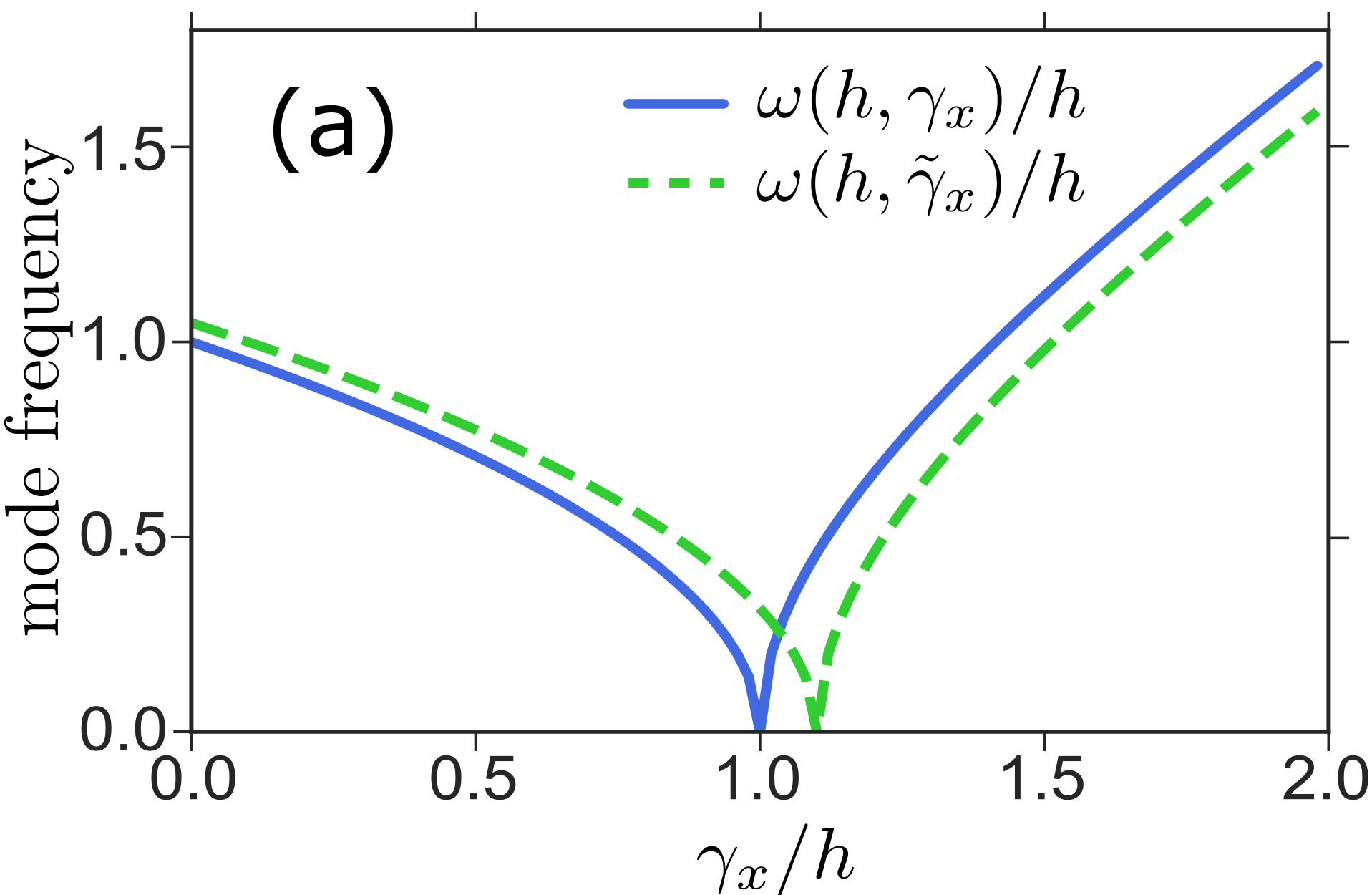}
\includegraphics[width=0.49 \columnwidth]{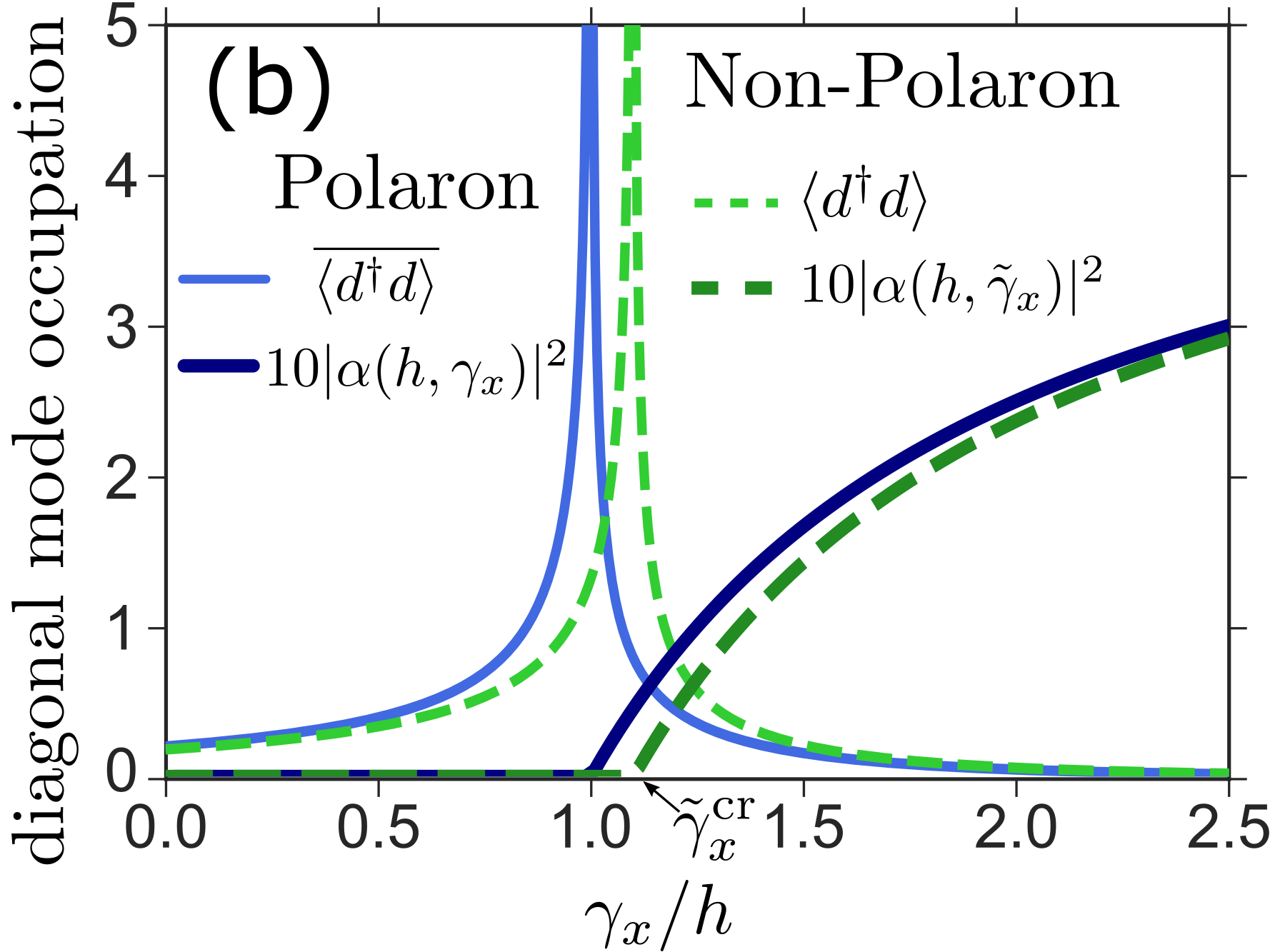}
\includegraphics[width=0.49 \columnwidth]{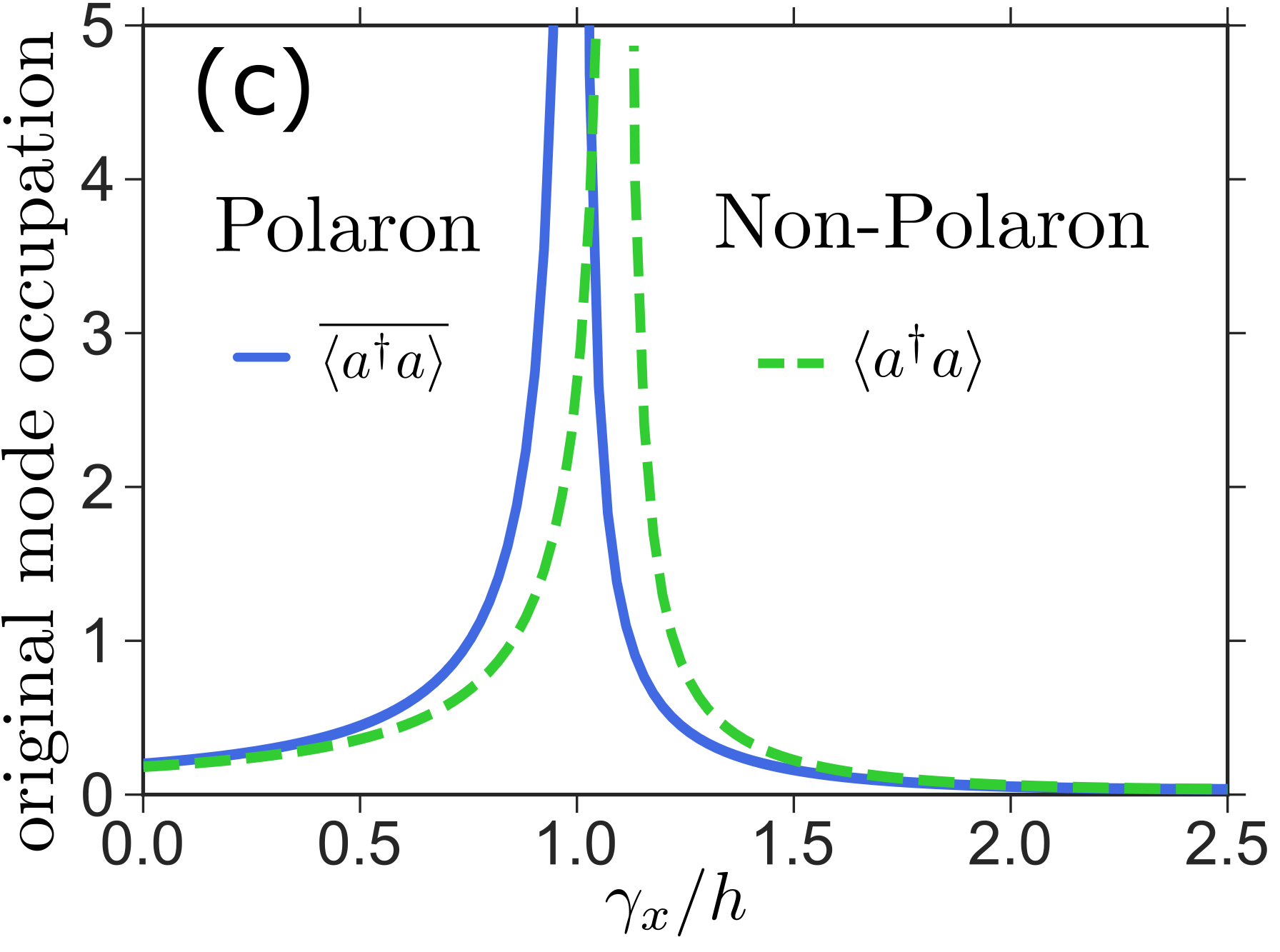}
\caption{(a) LMG oscillator frequency $\omega(h,\gamma_x)$ or $\omega(h,\tilde\gamma_x)$, (b) diagonal frame steady-state mode occupations $\overline{\avg{d^\dag d}}$  ($\avg{d^\dag d}$), (c) non-diagonal frame steady-state mode occupations $\overline{\avg{a^\dag a}} (\avg{a^\dag a})$  for the polaron (solid) and non-polaron (dashed) master equations. 
Divergent mode occupations indicate the position of the QPT where the excitation frequency vanishes. 
For the polaron treatment, the QPT position stays at $\gamma_x/h = 1$ just as in the isolated LMG model
in contrast to the shift predicted by the BMS master equation.  
Parameters: $\eta=2\pi \cdot 0.1, \omega_c = 0.5 h, \beta = 1.79/h$.
}
\label{fig:fluctuations} 
\end{figure}

The master equations appear simple only in a displaced and rotated frame. 
When transformed back, the steady-state populations
$\avg{d^\dagger d} = \trace{d^\dagger d \rho}$ and 
$\overline{\avg{d^\dagger d}} = \trace{d^\dagger d \bar\rho}$ actually measure displacements around the mean-field.
Fig.~\ref{fig:fluctuations} compares the occupation number and system frequency with (solid) and without (dashed) polaron transform.
Panel (a) demonstrates that the LMG energy gap is in the BMS treatment strongly modified by dissipation, such that in the vicinity of the closed QPT the non-polaron and polaron treatments lead to very different results.
Panel (b) shows the fluctuations in the diagonal basis $\overline{\avg{d^\dag d}}$ ($\avg{d^\dag d}$) around the mean-field $\alpha(h,\gamma_x)$ (or $\alpha(h,\tilde\gamma_x)$) in the polaron (or non-polaron) frame.
Finally, panel (c) shows the mode occupation 
$\avg{a^\dagger a} = \sinh^2(\varphi(h,\gamma_x)) + 2 \cosh^2(\varphi(h,\gamma_x)) \avg{d^\dagger d}$ (and analogous in the symmetry-broken phase) in the non-diagonal basis.
These are directly related to the deviations of the $J_z$-expectation value from its mean-field solution, compare App.~\ref{APP:tdlimit}.
Since the frequency $\omega(h,\tilde\gamma_x)$ (Tab.~\ref{tab:0}) vanishes at 
$\gamma_x = h + \frac{\eta \cdot \omega_c}{\pi}$ in the non-polaron frame, 
the BMS approximations break down around the original QPT position, see dashed line in Fig.~\ref{fig:fluctuations}(a).
Mode occupations in both the diagonal and non-diagonal bases diverge at the QPT point, see the dashed lines in Fig.~\ref{fig:fluctuations}(b-c). 
In particular, in the polaron frame the fluctuation divergence occurs around the original quantum critical point at $\gamma_x = h$, see the solid lines in Fig.~\ref{fig:fluctuations}.

\subsection{Waiting times}

The coupling to the reservoir does not only modify the system properties but may also lead to the emission or absorption of reservoir excitations (i.e., photons or phonons depending on the model implementation), which can in principle be measured independently.
Classifying these events into classes $\nu$ describing e.g.\ emissions or absorptions, the waiting-time distribution between two such system-bath exchange processes of type $\mu$ after $\nu$ is characterized by~\cite{brandes2008waiting} 
\begin{equation}
\label{eq:wt_definition}
\mathcal{w}_{\mu\nu}(\tau) = \frac{{\rm Tr}\left(\mathcal{J}_\mu \exp(\mathcal{L}_0 \tau)\mathcal{J}_\nu \rho\right)}{{\rm Tr} \left(\mathcal{J}_\nu \rho \right)}\,.
\end{equation} 
Here $\mathcal{J}_\mu, \mathcal{L}_0$ are super operators describing the jump $\mu$ and the no-jump evolution $\mathcal{L}_0$. 
For example, in master equation~\eqref{EQ:density_matrix}, there are only two distinct types of jumps, emission `e' and absorption `a'. 
Their corresponding super-operators are then acting as
\begin{align}
\mathcal{J}_{e} \rho &= F_e d \rho d^\dag\,,\qquad
\mathcal{J}_{a} \rho = F_a d^\dag \rho d\,, \notag\\
\mathcal{L}_0 \rho &= -\ii \left[\omega d^\dagger d, \rho\right] - \frac{F_e}{2} \left\{d^\dagger d,  \rho\right\}
- \frac{F_a}{2} \left\{d d^\dagger, \rho\right\}\,,
\end{align}
such that the total Liouvillian is decomposable as $\mathcal{L} = \mathcal{L}_0 + \mathcal{J}_e + \mathcal{J}_a$.
The same equations are valid in the polaron frame~\eqref{eq:density_matrix_polaron}, just with the corresponding overbar on the variables. 

It is straightforward to go to a frame where the Hamiltonian dynamics is absorbed $\tilde{\rho} = e^{+\ii \omega t d^\dagger d} \rho e^{-\ii \omega t d^\dagger d}$, we see that the whole Liouvillian in this frame $\tilde{\mathcal L}$ is just proportional to the spectral density, evaluated at the 
system transition frequency $\omega$.
Thereby, it enters as a single parameter, a different spectral density could be interpreted as a rescaling
$\Gamma(\omega) \to \alpha \Gamma(\omega)$, which would imply ${\cal L}_0 \to \alpha {\cal L}_0$ and ${\cal J}_\mu \to \alpha {\cal J}_\mu$.
These transformations would only lead to a trivial stretching of the waiting time distribution
$\mathcal{w}_{\mu\nu}(\tau) \to \alpha \mathcal{w}_{\mu\nu}(\alpha \tau)$, compare also Eq.~(\ref{EQ:waitingtime_explicit}).

Since the LMG Hamiltonian and the steady state~\eqref{eq:density_matrix_steady_state} are diagonal, analytic expressions for the waiting time distributions can be derived, see App.~\ref{APP:waiting_time}. 
\begin{figure}
\includegraphics[width=0.49 \columnwidth]{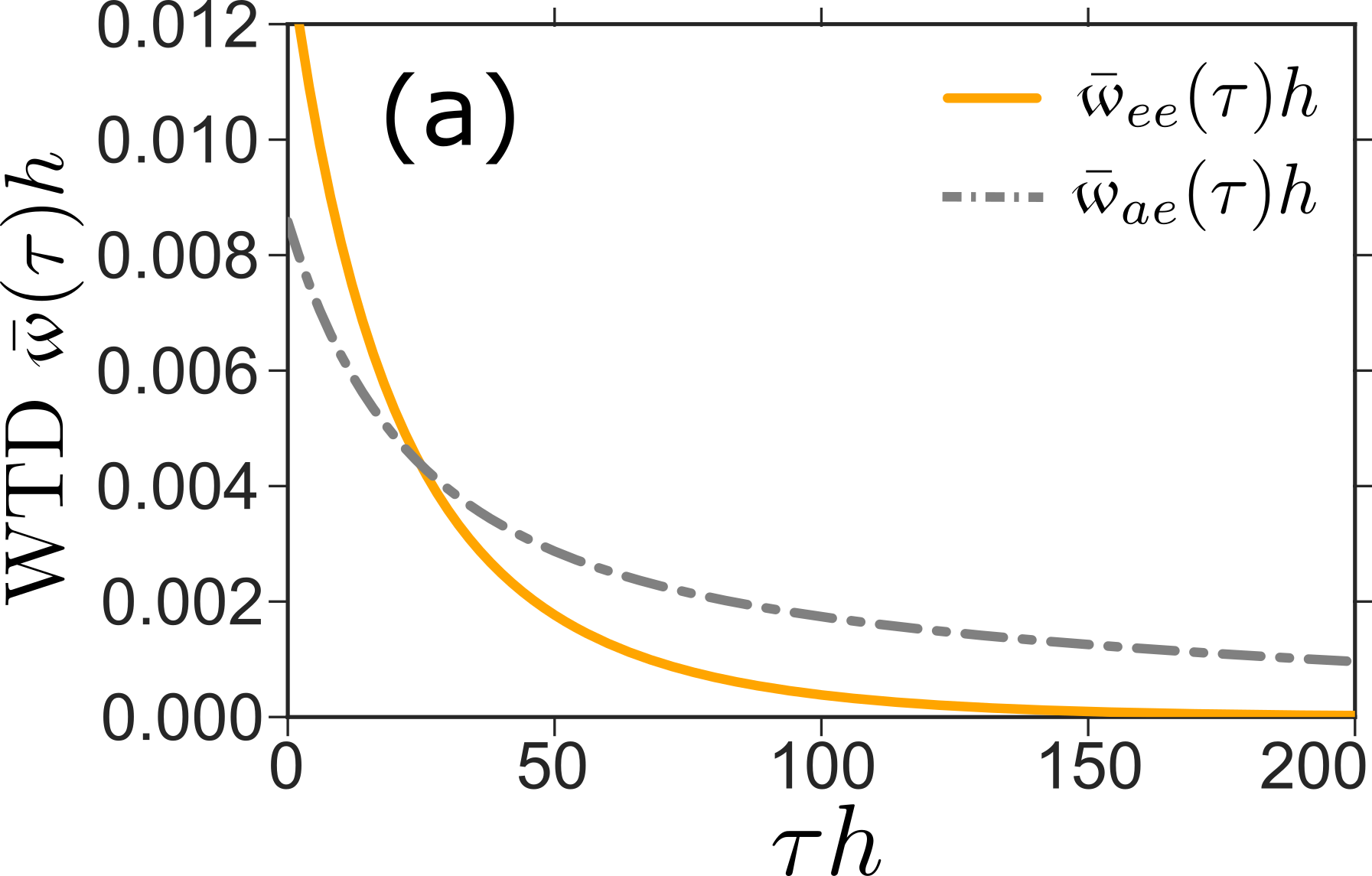}
\includegraphics[width=0.49 \columnwidth]{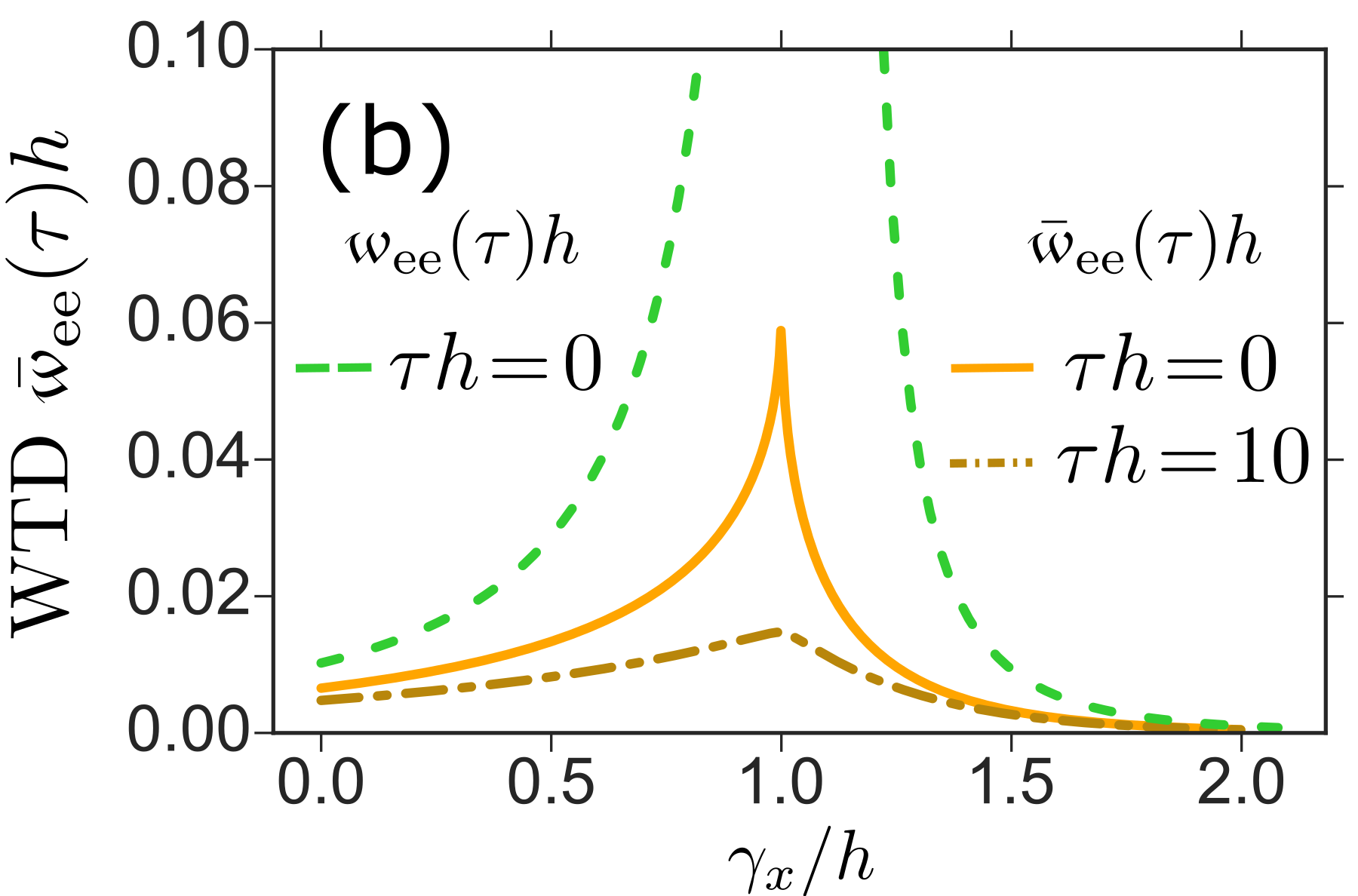}
\caption{Waiting time distributions (WTD) between two emission (absorption and emission) events $\bar{\mathcal{w}}_{ee(ae)}$ (solid, dot-dashed) calculated in the polaron frame as a function of $\tau$ (a) for a fixed 
$\gamma_x$ value and (b) distribution $\bar{\mathcal{w}}_{ee}$ as a function of $\gamma_x$ for two different fixed $\tau$ values (b). 
Additionally, the WTD in the non-polaron frame is shown in (b) for $\tau = 0$ case (dashed), which wrongly diverges around the shifted critical point. 
At the true critical point a non-analytic dependence of the distribution on the intra-spin coupling strength $\gamma_x$ is clearly visible, within the polaron treatment however all WTDs remain finite.
Parameters: $\eta=2\pi \cdot 0.1, \omega_c = 0.5 h, \beta = 1.79/h$, (a) $\gamma_x = 0.5 h$. 
}
\label{fig:waiting_times}
\end{figure}
In Fig.~\ref{fig:waiting_times} we show two waiting-time distributions $\bar{\mathcal{w}}_{ee(ae)}$ as a function of time $\tau$ for fixed coupling strength 
$\gamma_x$ (a) and the repeated-emission waiting-time distribution $\bar{\mathcal{w}}_{ee}(\tau)$ as a function of $\gamma_x$ for two fixed waiting times $\tau$ (b). 
A typical feature of a thermal state is bunching of emitted photons, which we see in Fig.~\ref{fig:waiting_times}(a): 
After an emission event the same event has the highest probability for $\tau \to 0$, thus immediately. 
When looking at waiting time distributions of different phases, like in panel (a), a significant difference is not visible.
However, fixing the waiting time $\tau$ and varying $\gamma_x$ we find, that the waiting times have their maximum at the position of QPT, see Fig.~\ref{fig:waiting_times}(b). 
Essentially, this is related to the divergence of $n_B(\omega)$ when the energy gap vanishes.
Whereas the non-polaron treatment predicts a divergence of waiting times around the critical point $\tilde{\gamma}_x^{\rm cr}$, see the dashed curve in Fig.~\ref{fig:waiting_times}(b), the waiting times within the polaron approach remain finite but depend non-analytically on the Hamiltonian parameters.

Therefore, the quantum-critical behaviour is not only reflected in system-intrinsic observables like mode occupations but also in reservoir observables like the statistics of photoemission events.

\section{Summary}

We have investigated the open LMG model by using a polaron transform technique 
that also allows us to address the vicinity of the critical point.

First, within the polaron treatment, we have found that the position of the QPT is robust when starting 
from an initial Hamiltonian with a lower spectral bound. 
This shows that the choice of the starting Hamiltonian should be discussed with care for critical models, even when 
treated as weakly coupled.

Second, whereas far from the QPT, the approach presented here reproduces naive master equation treatments, 
it remains also valid in the vicinity of the QPT.
In the transformed frame, the effective interaction
scaled with the energy gap of the system Hamiltonian, which admits a perturbative treatment at the critical point.
We therefore expect that the polaron-master equation approach is also applicable 
to other models that bilinearly couple to bosonic reservoirs via position operators.

Interestingly, we obtained that for a single reservoir the stationary properties are determined by those of the isolated system alone,
such that a standard analysis applies.

The critical behaviour (and its possible renormalization) can be detected with system observables like magnetization or mode occupations but is also visible in reservoir observables like waiting-time distributions, which remain finite in the polaron frame.
We hope that our study of the LMG model paves the way for further quantitative investigations of dissipative quantum-critical systems, 
e.g. by capturing higher eigenstates by augmented variational polaron treatments~\cite{mccutcheon2011a} or by investigating the non-equilibrium dynamics
of critical setups.


\section*{Acknowledgements}\vspace{-2mm}
The authors gratefully acknowledge financial support from the DFG (grants BR 1528/9-1, BR 1528/8-2, and SFB 910) as well as fruitful discussions with M. Kloc, A. Knorr, and C. W\"achtler.


\appendix
 
\section{Thermodynamic limit of large spin operators}\label{APP:tdlimit}

Without any displacement, the Holstein-Primakoff representation leads to a simple large-$N$ expansion 
\begin{align}
J_- &\approx \sqrt{N} b^\dag\,,\qquad
J_+  \approx \sqrt{N} b\,,\notag\\
J_z &= \frac{N}{2} - b^\dag b\,,
\end{align} 
where we have neglected terms that vanish in the thermodynamic limit.
Insertion of these approximations lead to the Hamiltonians for the normal phase, and in effect, no term of order $\sqrt{N}$ occurs in the Hamiltonian.

In the symmetry-broken phase, one allows for a displacement $b = a + \alpha \sqrt{N}$ with bosonic operators $a$ and in general complex number $\alpha$.
Then, the large-$N$ expansion of the large spin operators is more complicated
\begin{align}
J_- &\approx N \alpha^* \sqrt{1-\abs{\alpha}^2}\notag\\
&\qquad+ \sqrt{N} \sqrt{1-\abs{\alpha}^2} \left[a^\dagger - \frac{1}{2} \frac{(\alpha^*)^2 a + \abs{\alpha}^2 a^\dagger}{1-\abs{\alpha}^2}\right]\notag\\
&\qquad- \frac{\sqrt{1-\abs{\alpha}^2}}{2\left(1-\abs{\alpha}^2\right)}\Big[
\alpha (a^\dagger)^2 + 2 \alpha^* a^\dagger a\notag\\
&\qquad\qquad\qquad+\frac{\alpha^* \left(\alpha^* a + \alpha a^\dagger\right)^2}{4 \left(1-\abs{\alpha}^2\right)}\Big]\,,\notag\\
J_+ &\approx N \alpha \sqrt{1-\abs{\alpha}^2}\notag\\
&\qquad+ \sqrt{N} \sqrt{1-\abs{\alpha}^2} \left[a - \frac{1}{2} \frac{\alpha^2 a^\dagger + \abs{\alpha}^2 a}{1-\abs{\alpha}^2}\right]\notag\\
&\qquad- \frac{\sqrt{1-\abs{\alpha}^2}}{2\left(1-\abs{\alpha}^2\right)}\Big[
\alpha^* a^2 + 2 \alpha a^\dagger a\notag\\
&\qquad\qquad\qquad+\frac{\alpha \left(\alpha^* a + \alpha a^\dagger\right)^2}{4 \left(1-\abs{\alpha}^2\right)}\Big]
\,,\notag\\
J_z &= N \left(\frac{1}{2}-\abs{\alpha}^2\right) - \sqrt{N} \left(\alpha^* a + \alpha a^\dag\right) - a^\dag a\,.
\end{align}
For consistency, one can check that by setting $\alpha\to 0$, the previous representation is reproduced.
Insertion of this expansion leads to the Hamiltonians for the symmetry-broken phase, and the displacement $\alpha$ is chosen such 
that the $\sqrt{N}$ terms in the LMG Hamiltonian vanish.
One might be tempted to neglect the last expansion terms in $J_\pm$ from the beginning, as these operators enter the Hamiltonian always with 
a factor of $1/\sqrt{N}$.
However, we stress that in terms like $J_x^2/N$ they will yield a non-vanishing contribution and thus need to be considered to obtain the correct 
spectra of the LMG model.

\section{Polaron transform}
\label{app:polaron_derivation}

Here we provide more details how to derive Eq.~\eqref{eq:H_LMG_And_Bath_Polaron} in the main text. 
Using the Hadamard lemma
\begin{align}
e^{+X} Y e^{-X} &= \sum_{m=0}^\infty \frac{1}{m!} \left[X,Y\right]_m\,,\\
\left[X,Y\right]_m &= \left[X,\left[X,Y\right]_{m-1}\right]\,,\qquad
[X,Y]_0=Y\,,\notag
\end{align}
one can see that the polaron transform~(\ref{eq:Polaron_trafo}) leads to 
\begin{align}
U_p^\dag c_k U_p &= c_k - \frac{J_x}{\sqrt{N}} \frac{g_k}{\nu_k}\,,
\end{align}
and analogous for the transformation of the creation operator.
Furthermore, it is trivial to see that $U_p^\dag J_x U_p = J_x$.
From this, it directly follows that the polaron-transform of the interaction and reservoir Hamiltonian becomes
\begin{align}
\label{eq:help_a}
U_p^\dag \rb{c_k^\dag + \frac{g_k}{\sqrt{N} \nu_k}J_x}\rb{c_k + \frac{g_k}{\sqrt{N} \nu_k}J_x} U_p  
&= c_k^\dag c_k\,.
\end{align}

In addition, the polaron transform of $J_z$ has to be calculated, which yields via the commutation relations $[J_x,J_y]=\ii J_z$ the relation
\begin{align}
\label{eq:jz_polaron_non_polaron_connection}
U^\dag_p J_z U_p &= J_z \cosh(\hat{B}) - \ii J_y \sinh(\hat{B})\,, 
\end{align}
where $\hat{B}$ is defined in~\eqref{eq:Polaron_trafo} in the main text.

Therefore, the full polaron-transformed Hamiltonian $H_{\rm tot}$ becomes
\begin{align}
\label{eq:H_LMG_And_Bath_ini_polaron}
U_p^\dag H_{tot} U_p 
 &= -h D J_z - \frac{\gamma_x}{N}
J_x^2 + \sum_k \nu_k b_k^\dag b_k\notag \\
 &\quad- h \cdot \left[J_z \cdot \left(\cosh(\hat{B})-D\right) - \ii J_y \sinh(\hat{B})\right]\,,\notag
\end{align}
such that there is no rescaling of the spin-spin interaction $\gamma_x$. 
We have also already inserted the temperature-dependent shift $D$, which is necessary in order to ensure that the first order expectation values of the system-reservoir coupling operators vanish, eventually yielding Eq.~\eqref{eq:H_LMG_And_Bath_Polaron} in the main text.
For the $\sinh$-term this is not necessary as its expectation value vanishes anyhow.

\section{Magnetization}\label{APP:magnetization}

It is well known that for a Hamiltonian depending on an external parameter $\lambda$ (which for your model could be $h$ or $\gamma_x$), the canonical partition function
\begin{align}
Z=\trace{e^{-\beta H(\lambda)}}
\end{align}
allows to evaluate the thermal expectation value of particular operators
\begin{align}
\frac{-1}{\beta} \frac{\partial \ln Z}{\partial \lambda} &= \frac{-1}{Z \beta} \trace{\frac{\partial}{\partial \lambda} e^{-\beta H(\lambda)}}\notag\\
&= \frac{1}{Z} \sum_{n=1}^\infty \frac{(-\beta)^{n-1}}{(n-1)!} \trace{\frac{\partial H(\lambda)}{\partial \lambda} H^{n-1}(\lambda)}\notag\\
&= \frac{1}{Z} \trace{\frac{\partial H(\lambda)}{\partial \lambda} e^{-\beta H(\lambda)}}
= \avg{\frac{\partial H(\lambda)}{\partial \lambda}}\,,
\end{align}
where we have used the invariance of the trace under cyclic permutations to sort all derivatives of $H(\lambda)$ to the left.

In particular, for a harmonic oscillator $H=E_0(\lambda)+\omega(\lambda) a^\dagger(\lambda) a(\lambda)$ with bosonic operators $a(\lambda)$, the partition function becomes
\begin{align}
Z = \frac{e^{-\beta E_0(\lambda)}}{1-e^{-\beta \omega(\lambda)}}\,.
\end{align}
With $\lambda\to -h$, this eventually leads to Eq.~(\ref{EQ:mag_ss}) in the main text.

\section{Waiting time distribution}\label{APP:waiting_time}

Starting from the spectral decomposition of a thermal state in terms of Fock states
\begin{align}
\rho &= \frac{e^{-\beta \omega d^\dagger d}}{\trace{e^{-\beta \omega d^\dagger d}}} = \sum_{n=0}^\infty P_n \ket{n}\bra{n}\notag\\
P_n &= \left(\frac{n_B}{1+n_B}\right)^n \frac{1}{1+n_B}\,,
\end{align}
with the shorthand notation $n_B = [e^{\beta \omega}-1]^{-1}$,
it is straightforward to compute the action of the emission or absorption jump superoperators 
\begin{align}
\mathcal{J}_e \rho &= F_e \sum_{n=0}^\infty P_{n+1} (n+1) \ket{n}\bra{n}\,,\notag\\
\mathcal{J}_a \rho &= F_a \sum_{n=1}^\infty P_{n-1} n \ket{n}\bra{n}\,,
\end{align}
which also implies
\begin{align}
\trace{\mathcal{J}_e \rho} = \trace{\mathcal{J}_a \rho} = \Gamma n_B (1+n_B)\,,
\end{align}
where $\Gamma=A^2(h,\tilde\gamma_x) \Gamma(\omega(h,\tilde\gamma_x))$ or $\Gamma = \bar{A}^2(h,\gamma_x) \bar{\Gamma}(\omega(h,\gamma_x))$ in the main text.
Since $\mathcal{L}_0$ does not induce transitions between different Fock states, its action on a diagonal density matrix can be computed via
\begin{align}
e^{\mathcal{L}_0 t} \ket{n}\bra{n} = e^{-\left[(1+n_B) n + n_B (1+n)\right] \Gamma t} \ket{n}\bra{n}\,,
\end{align}
which implies for the relevant terms
\begin{align}\label{EQ:waitingtime_explicit}
\mathcal{w}_{ee}(\tau) &= \frac{2 \Gamma n_B (1+n_B) e^{(2+3 n_B) \Gamma \tau}}{\left[(1+n_B) e^{(1+2 n_B)\Gamma \tau} - n_B\right]^3}\,,\notag\\
\mathcal{w}_{ae}(\tau) &= \frac{\Gamma n_B e^{(2+3 n_B) \Gamma \tau} \left[n_B+ (1+n_B) e^{(1+2 n_B) \Gamma \tau}\right]}{\left[(1+n_B) e^{(1+2 n_B)\Gamma \tau} - n_B\right]^3}\,,\notag\\
\mathcal{w}_{ea}(\tau) &= \frac{\Gamma (1+n_B) e^{(1+n_B) \Gamma \tau} \left[n_B+ (1+n_B) e^{(1+2 n_B) \Gamma \tau}\right]}{\left[(1+n_B) e^{(1+2 n_B)\Gamma \tau} - n_B\right]^3}\,,\notag\\
\mathcal{w}_{aa}(\tau) &= \frac{2 \Gamma n_B (1+n_B) e^{(2+3 n_B) \Gamma \tau}}{\left[(1+n_B) e^{(1+2 n_B)\Gamma \tau} - n_B\right]^3}\,.
\end{align}
For consistency, we note that the normalization conditions 
$\int \left(\mathcal{w}_{ae}(\tau) + \mathcal{w}_{ee}(\tau)\right)d\tau = 1$ and 
$\int \left(\mathcal{w}_{aa}(\tau) + \mathcal{w}_{ea}(\tau)\right)d\tau = 1$ always hold, which simply reflects the 
fact that only emission or absorption processes can occur.
Furthermore, in the low-temperature limit $n_B \to 0$, only the conditional waiting time distribution for emission after absorption can survive 
$\mathcal{w}_{ea} \to \Gamma e^{-\Gamma \tau}$: Once a photon has been absorbed from the reservoir, it must be emitted again since no further absorption is likely to occur.
For $\tau \gg 1$ all waiting time distributions $\bar{\mathcal{w}}_{\mu\nu}$ decay to zero.


\begin{thebibliography}{83}%
	\makeatletter
	\providecommand \@ifxundefined [1]{%
		\@ifx{#1\undefined}
	}%
	\providecommand \@ifnum [1]{%
		\ifnum #1\expandafter \@firstoftwo
		\else \expandafter \@secondoftwo
		\fi
	}%
	\providecommand \@ifx [1]{%
		\ifx #1\expandafter \@firstoftwo
		\else \expandafter \@secondoftwo
		\fi
	}%
	\providecommand \natexlab [1]{#1}%
	\providecommand \enquote  [1]{``#1''}%
	\providecommand \bibnamefont  [1]{#1}%
	\providecommand \bibfnamefont [1]{#1}%
	\providecommand \citenamefont [1]{#1}%
	\providecommand \href@noop [0]{\@secondoftwo}%
	\providecommand \href [0]{\begingroup \@sanitize@url \@href}%
	\providecommand \@href[1]{\@@startlink{#1}\@@href}%
	\providecommand \@@href[1]{\endgroup#1\@@endlink}%
	\providecommand \@sanitize@url [0]{\catcode `\\12\catcode `\$12\catcode
		`\&12\catcode `\#12\catcode `\^12\catcode `\_12\catcode `\%12\relax}%
	\providecommand \@@startlink[1]{}%
	\providecommand \@@endlink[0]{}%
	\providecommand \url  [0]{\begingroup\@sanitize@url \@url }%
	\providecommand \@url [1]{\endgroup\@href {#1}{\urlprefix }}%
	\providecommand \urlprefix  [0]{URL }%
	\providecommand \Eprint [0]{\href }%
	\providecommand \doibase [0]{http://dx.doi.org/}%
	\providecommand \selectlanguage [0]{\@gobble}%
	\providecommand \bibinfo  [0]{\@secondoftwo}%
	\providecommand \bibfield  [0]{\@secondoftwo}%
	\providecommand \translation [1]{[#1]}%
	\providecommand \BibitemOpen [0]{}%
	\providecommand \bibitemStop [0]{}%
	\providecommand \bibitemNoStop [0]{.\EOS\space}%
	\providecommand \EOS [0]{\spacefactor3000\relax}%
	\providecommand \BibitemShut  [1]{\csname bibitem#1\endcsname}%
	\let\auto@bib@innerbib\@empty
	\bibitem [{\citenamefont {Sachdev}(2007)}]{Sachdev-QPT}%
	\BibitemOpen
	\bibfield  {author} {\bibinfo {author} {\bibfnamefont {S.}~\bibnamefont
			{Sachdev}},\ }\href@noop {} {\emph {\bibinfo {title} {Quantum phase
				transitions}}}\ (\bibinfo  {publisher} {Wiley Online Library},\ \bibinfo
	{year} {2007})\BibitemShut {NoStop}%
	\bibitem [{\citenamefont {Ribeiro}\ \emph {et~al.}(2007)\citenamefont
		{Ribeiro}, \citenamefont {Vidal},\ and\ \citenamefont
		{Mosseri}}]{LMG-thermodynamical_limit-Mosseri}%
	\BibitemOpen
	\bibfield  {author} {\bibinfo {author} {\bibfnamefont {P.}~\bibnamefont
			{Ribeiro}}, \bibinfo {author} {\bibfnamefont {J.}~\bibnamefont {Vidal}}, \
		and\ \bibinfo {author} {\bibfnamefont {R.}~\bibnamefont {Mosseri}},\ }\href
	{\doibase 10.1103/PhysRevLett.99.050402} {\bibfield  {journal} {\bibinfo
			{journal} {Phys. Rev. Lett.}\ }\textbf {\bibinfo {volume} {99}},\ \bibinfo
		{pages} {050402} (\bibinfo {year} {2007})}\BibitemShut {NoStop}%
	\bibitem [{\citenamefont {Bastarrachea-Magnani}\ \emph
		{et~al.}(2014)\citenamefont {Bastarrachea-Magnani}, \citenamefont
		{Lerma-Hern\'andez},\ and\ \citenamefont
		{Hirsch}}]{Hirsch-Dicke_TC_-quantum-and-semi-analysis-chaos}%
	\BibitemOpen
	\bibfield  {author} {\bibinfo {author} {\bibfnamefont {M.~A.}\ \bibnamefont
			{Bastarrachea-Magnani}}, \bibinfo {author} {\bibfnamefont {S.}~\bibnamefont
			{Lerma-Hern\'andez}}, \ and\ \bibinfo {author} {\bibfnamefont {J.~G.}\
			\bibnamefont {Hirsch}},\ }\href {\doibase 10.1103/PhysRevA.89.032102}
	{\bibfield  {journal} {\bibinfo  {journal} {Phys. Rev. A}\ }\textbf {\bibinfo
			{volume} {89}},\ \bibinfo {pages} {032102} (\bibinfo {year}
		{2014})}\BibitemShut {NoStop}%
	\bibitem [{\citenamefont {Lambert}\ \emph {et~al.}(2004)\citenamefont
		{Lambert}, \citenamefont {Emary},\ and\ \citenamefont
		{Brandes}}]{Dicke_Entanglement_and_QPT-Brandes}%
	\BibitemOpen
	\bibfield  {author} {\bibinfo {author} {\bibfnamefont {N.}~\bibnamefont
			{Lambert}}, \bibinfo {author} {\bibfnamefont {C.}~\bibnamefont {Emary}}, \
		and\ \bibinfo {author} {\bibfnamefont {T.}~\bibnamefont {Brandes}},\ }\href
	{\doibase 10.1103/PhysRevLett.92.073602} {\bibfield  {journal} {\bibinfo
			{journal} {Phys. Rev. Lett.}\ }\textbf {\bibinfo {volume} {92}},\ \bibinfo
		{pages} {073602} (\bibinfo {year} {2004})}\BibitemShut {NoStop}%
	\bibitem [{\citenamefont {Baumann}\ \emph {et~al.}(2010)\citenamefont
		{Baumann}, \citenamefont {Guerlin}, \citenamefont {Brennecke},\ and\
		\citenamefont {Esslinger}}]{Baumann-Dicke_qpt}%
	\BibitemOpen
	\bibfield  {author} {\bibinfo {author} {\bibfnamefont {K.}~\bibnamefont
			{Baumann}}, \bibinfo {author} {\bibfnamefont {C.}~\bibnamefont {Guerlin}},
		\bibinfo {author} {\bibfnamefont {F.}~\bibnamefont {Brennecke}}, \ and\
		\bibinfo {author} {\bibfnamefont {T.}~\bibnamefont {Esslinger}},\ }\href@noop
	{} {\bibfield  {journal} {\bibinfo  {journal} {Nature (London)}\ }\textbf
		{\bibinfo {volume} {464}} (\bibinfo {year} {2010})}\BibitemShut {NoStop}%
	\bibitem [{\citenamefont {Baumann}\ \emph {et~al.}(2011)\citenamefont
		{Baumann}, \citenamefont {Mottl}, \citenamefont {Brennecke},\ and\
		\citenamefont {Esslinger}}]{Baumann-symm-break-in-Dicke-QPT}%
	\BibitemOpen
	\bibfield  {author} {\bibinfo {author} {\bibfnamefont {K.}~\bibnamefont
			{Baumann}}, \bibinfo {author} {\bibfnamefont {R.}~\bibnamefont {Mottl}},
		\bibinfo {author} {\bibfnamefont {F.}~\bibnamefont {Brennecke}}, \ and\
		\bibinfo {author} {\bibfnamefont {T.}~\bibnamefont {Esslinger}},\ }\href
	{\doibase 10.1103/PhysRevLett.107.140402} {\bibfield  {journal} {\bibinfo
			{journal} {Phys. Rev. Lett.}\ }\textbf {\bibinfo {volume} {107}},\ \bibinfo
		{pages} {140402} (\bibinfo {year} {2011})}\BibitemShut {NoStop}%
	\bibitem [{\citenamefont {Brennecke}\ \emph {et~al.}(2013)\citenamefont
		{Brennecke}, \citenamefont {Mottl}, \citenamefont {Baumann}, \citenamefont
		{Landig}, \citenamefont {Donner},\ and\ \citenamefont
		{Esslinger}}]{brennecke2013real}%
	\BibitemOpen
	\bibfield  {author} {\bibinfo {author} {\bibfnamefont {F.}~\bibnamefont
			{Brennecke}}, \bibinfo {author} {\bibfnamefont {R.}~\bibnamefont {Mottl}},
		\bibinfo {author} {\bibfnamefont {K.}~\bibnamefont {Baumann}}, \bibinfo
		{author} {\bibfnamefont {R.}~\bibnamefont {Landig}}, \bibinfo {author}
		{\bibfnamefont {T.}~\bibnamefont {Donner}}, \ and\ \bibinfo {author}
		{\bibfnamefont {T.}~\bibnamefont {Esslinger}},\ }\href@noop {} {\bibfield
		{journal} {\bibinfo  {journal} {PNAS}\ }\textbf {\bibinfo {volume} {110 No.
				29}},\ \bibinfo {pages} {11763} (\bibinfo {year} {2013})}\BibitemShut
	{NoStop}%
	\bibitem [{\citenamefont {Zibold}\ \emph {et~al.}(2010)\citenamefont {Zibold},
		\citenamefont {Nicklas}, \citenamefont {Gross},\ and\ \citenamefont
		{Oberthaler}}]{LMG-Exp_Bifurcation_rabi_to_jesophson-Oberthaler}%
	\BibitemOpen
	\bibfield  {author} {\bibinfo {author} {\bibfnamefont {T.}~\bibnamefont
			{Zibold}}, \bibinfo {author} {\bibfnamefont {E.}~\bibnamefont {Nicklas}},
		\bibinfo {author} {\bibfnamefont {C.}~\bibnamefont {Gross}}, \ and\ \bibinfo
		{author} {\bibfnamefont {M.~K.}\ \bibnamefont {Oberthaler}},\ }\href
	{\doibase 10.1103/PhysRevLett.105.204101} {\bibfield  {journal} {\bibinfo
			{journal} {Phys. Rev. Lett.}\ }\textbf {\bibinfo {volume} {105}},\ \bibinfo
		{pages} {204101} (\bibinfo {year} {2010})}\BibitemShut {NoStop}%
	\bibitem [{\citenamefont {Ritsch}\ \emph {et~al.}(2013)\citenamefont {Ritsch},
		\citenamefont {Domokos}, \citenamefont {Brennecke},\ and\ \citenamefont
		{Esslinger}}]{Ritsch-Domokos-Cold-atoms-in-opt-potential}%
	\BibitemOpen
	\bibfield  {author} {\bibinfo {author} {\bibfnamefont {H.}~\bibnamefont
			{Ritsch}}, \bibinfo {author} {\bibfnamefont {P.}~\bibnamefont {Domokos}},
		\bibinfo {author} {\bibfnamefont {F.}~\bibnamefont {Brennecke}}, \ and\
		\bibinfo {author} {\bibfnamefont {T.}~\bibnamefont {Esslinger}},\ }\href
	{\doibase 10.1103/RevModPhys.85.553} {\bibfield  {journal} {\bibinfo
			{journal} {Rev. Mod. Phys.}\ }\textbf {\bibinfo {volume} {85}},\ \bibinfo
		{pages} {553} (\bibinfo {year} {2013})}\BibitemShut {NoStop}%
	\bibitem [{\citenamefont {Dicke}(1954)}]{Dicke-Dicke_Modell}%
	\BibitemOpen
	\bibfield  {author} {\bibinfo {author} {\bibfnamefont {R.~H.}\ \bibnamefont
			{Dicke}},\ }\href {\doibase 10.1103/PhysRev.93.99} {\bibfield  {journal}
		{\bibinfo  {journal} {Phys. Rev.}\ }\textbf {\bibinfo {volume} {93}},\
		\bibinfo {pages} {99} (\bibinfo {year} {1954})}\BibitemShut {NoStop}%
	\bibitem [{\citenamefont {Fusco}\ \emph {et~al.}(2016)\citenamefont {Fusco},
		\citenamefont {Paternostro},\ and\ \citenamefont {De~Chiara}}]{fusco2016a}%
	\BibitemOpen
	\bibfield  {author} {\bibinfo {author} {\bibfnamefont {L.}~\bibnamefont
			{Fusco}}, \bibinfo {author} {\bibfnamefont {M.}~\bibnamefont {Paternostro}},
		\ and\ \bibinfo {author} {\bibfnamefont {G.}~\bibnamefont {De~Chiara}},\
	}\href {\doibase 10.1103/PhysRevE.94.052122} {\bibfield  {journal} {\bibinfo
			{journal} {Physical Review E}\ }\textbf {\bibinfo {volume} {94}},\ \bibinfo
		{pages} {052122} (\bibinfo {year} {2016})}\BibitemShut {NoStop}%
	\bibitem [{\citenamefont {{\c{C}}akmak}\ \emph {et~al.}(2016)\citenamefont
		{{\c{C}}akmak}, \citenamefont {Altintas},\ and\ \citenamefont
		{E.~M{\"u}stecapl{\i}o{\u{g}}lu}}]{cakmak2016a}%
	\BibitemOpen
	\bibfield  {author} {\bibinfo {author} {\bibfnamefont {S.}~\bibnamefont
			{{\c{C}}akmak}}, \bibinfo {author} {\bibfnamefont {F.}~\bibnamefont
			{Altintas}}, \ and\ \bibinfo {author} {\bibfnamefont {{\"O}.}~\bibnamefont
			{E.~M{\"u}stecapl{\i}o{\u{g}}lu}},\ }\href {\doibase
		10.1140/epjp/i2016-16197-0} {\bibfield  {journal} {\bibinfo  {journal} {The
				European Physical Journal Plus}\ }\textbf {\bibinfo {volume} {131}},\
		\bibinfo {pages} {197} (\bibinfo {year} {2016})}\BibitemShut {NoStop}%
	\bibitem [{\citenamefont {Ma}\ \emph {et~al.}(2017)\citenamefont {Ma},
		\citenamefont {Su},\ and\ \citenamefont {Sun}}]{ma2017a}%
	\BibitemOpen
	\bibfield  {author} {\bibinfo {author} {\bibfnamefont {Y.-H.}\ \bibnamefont
			{Ma}}, \bibinfo {author} {\bibfnamefont {S.-H.}\ \bibnamefont {Su}}, \ and\
		\bibinfo {author} {\bibfnamefont {C.-P.}\ \bibnamefont {Sun}},\ }\href
	{\doibase 10.1103/PhysRevE.96.022143} {\bibfield  {journal} {\bibinfo
			{journal} {Phys. Rev. E}\ }\textbf {\bibinfo {volume} {96}},\ \bibinfo
		{pages} {022143} (\bibinfo {year} {2017})}\BibitemShut {NoStop}%
	\bibitem [{\citenamefont {Kloc}\ \emph {et~al.}()\citenamefont {Kloc},
		\citenamefont {Cejnar},\ and\ \citenamefont {Schaller}}]{kloc2019a}%
	\BibitemOpen
	\bibfield  {author} {\bibinfo {author} {\bibfnamefont {M.}~\bibnamefont
			{Kloc}}, \bibinfo {author} {\bibfnamefont {P.}~\bibnamefont {Cejnar}}, \ and\
		\bibinfo {author} {\bibfnamefont {G.}~\bibnamefont {Schaller}},\ }\href@noop
	{} {\bibfield  {journal} {\bibinfo  {journal} {Phys. Rev. E}\ }\textbf
		{\bibinfo {volume} {100}},\ \bibinfo {pages} {042126}}\BibitemShut {NoStop}%
	\bibitem [{\citenamefont {Bastidas}\ \emph {et~al.}(2012)\citenamefont
		{Bastidas}, \citenamefont {Emary}, \citenamefont {Regler},\ and\
		\citenamefont {Brandes}}]{Dicke-nonequilibrium_qpt-bastidas}%
	\BibitemOpen
	\bibfield  {author} {\bibinfo {author} {\bibfnamefont {V.~M.}\ \bibnamefont
			{Bastidas}}, \bibinfo {author} {\bibfnamefont {C.}~\bibnamefont {Emary}},
		\bibinfo {author} {\bibfnamefont {B.}~\bibnamefont {Regler}}, \ and\ \bibinfo
		{author} {\bibfnamefont {T.}~\bibnamefont {Brandes}},\ }\href {\doibase
		10.1103/PhysRevLett.108.043003} {\bibfield  {journal} {\bibinfo  {journal}
			{Phys. Rev. Lett.}\ }\textbf {\bibinfo {volume} {108}},\ \bibinfo {pages}
		{043003} (\bibinfo {year} {2012})}\BibitemShut {NoStop}%
	\bibitem [{\citenamefont {Engelhardt}\ \emph {et~al.}(2013)\citenamefont
		{Engelhardt}, \citenamefont {Bastidas}, \citenamefont {Emary},\ and\
		\citenamefont {Brandes}}]{LMG-ac_driven_QPT-Georg}%
	\BibitemOpen
	\bibfield  {author} {\bibinfo {author} {\bibfnamefont {G.}~\bibnamefont
			{Engelhardt}}, \bibinfo {author} {\bibfnamefont {V.~M.}\ \bibnamefont
			{Bastidas}}, \bibinfo {author} {\bibfnamefont {C.}~\bibnamefont {Emary}}, \
		and\ \bibinfo {author} {\bibfnamefont {T.}~\bibnamefont {Brandes}},\ }\href
	{\doibase 10.1103/PhysRevE.87.052110} {\bibfield  {journal} {\bibinfo
			{journal} {Phys. Rev. E}\ }\textbf {\bibinfo {volume} {87}},\ \bibinfo
		{pages} {052110} (\bibinfo {year} {2013})}\BibitemShut {NoStop}%
	\bibitem [{\citenamefont {Bastidas}\ \emph {et~al.}(2014)\citenamefont
		{Bastidas}, \citenamefont {Engelhardt}, \citenamefont {P\'erez-Fern\'andez},
		\citenamefont {Vogl},\ and\ \citenamefont
		{Brandes}}]{Bastidas-Critical_quasienergy_in_driven_systems}%
	\BibitemOpen
	\bibfield  {author} {\bibinfo {author} {\bibfnamefont {V.~M.}\ \bibnamefont
			{Bastidas}}, \bibinfo {author} {\bibfnamefont {G.}~\bibnamefont
			{Engelhardt}}, \bibinfo {author} {\bibfnamefont {P.}~\bibnamefont
			{P\'erez-Fern\'andez}}, \bibinfo {author} {\bibfnamefont {M.}~\bibnamefont
			{Vogl}}, \ and\ \bibinfo {author} {\bibfnamefont {T.}~\bibnamefont
			{Brandes}},\ }\href {\doibase 10.1103/PhysRevA.90.063628} {\bibfield
		{journal} {\bibinfo  {journal} {Phys. Rev. A}\ }\textbf {\bibinfo {volume}
			{90}},\ \bibinfo {pages} {063628} (\bibinfo {year} {2014})}\BibitemShut
	{NoStop}%
	\bibitem [{\citenamefont {Acevedo}\ \emph {et~al.}(2015)\citenamefont
		{Acevedo}, \citenamefont {Quiroga}, \citenamefont {Rodr{\'\i}guez},\ and\
		\citenamefont
		{Johnson}}]{Dicke-Robust_quantum_correlation_with_linear_increased_coupling-Acevedo}%
	\BibitemOpen
	\bibfield  {author} {\bibinfo {author} {\bibfnamefont {O.}~\bibnamefont
			{Acevedo}}, \bibinfo {author} {\bibfnamefont {L.}~\bibnamefont {Quiroga}},
		\bibinfo {author} {\bibfnamefont {F.}~\bibnamefont {Rodr{\'\i}guez}}, \ and\
		\bibinfo {author} {\bibfnamefont {N.}~\bibnamefont {Johnson}},\ }\href@noop
	{} {\bibfield  {journal} {\bibinfo  {journal} {New Journal of Physics}\
		}\textbf {\bibinfo {volume} {17}},\ \bibinfo {pages} {093005} (\bibinfo
		{year} {2015})}\BibitemShut {NoStop}%
	\bibitem [{\citenamefont {Kopylov}\ \emph {et~al.}(2017)\citenamefont
		{Kopylov}, \citenamefont {Schaller},\ and\ \citenamefont
		{Brandes}}]{LMG-Nonadiabatic_dynamics_of_ESQPT-kopylov}%
	\BibitemOpen
	\bibfield  {author} {\bibinfo {author} {\bibfnamefont {W.}~\bibnamefont
			{Kopylov}}, \bibinfo {author} {\bibfnamefont {G.}~\bibnamefont {Schaller}}, \
		and\ \bibinfo {author} {\bibfnamefont {T.}~\bibnamefont {Brandes}},\ }\href
	{\doibase 10.1103/PhysRevE.96.012153} {\bibfield  {journal} {\bibinfo
			{journal} {Phys. Rev. E}\ }\textbf {\bibinfo {volume} {96}},\ \bibinfo
		{pages} {012153} (\bibinfo {year} {2017})}\BibitemShut {NoStop}%
	\bibitem [{\citenamefont
		{Campbell}(2016)}]{LMG-Criticality_revealed_and_quench_dynamics-Campbell}%
	\BibitemOpen
	\bibfield  {author} {\bibinfo {author} {\bibfnamefont {S.}~\bibnamefont
			{Campbell}},\ }\href {\doibase 10.1103/PhysRevB.94.184403} {\bibfield
		{journal} {\bibinfo  {journal} {Phys. Rev. B}\ }\textbf {\bibinfo {volume}
			{94}},\ \bibinfo {pages} {184403} (\bibinfo {year} {2016})}\BibitemShut
	{NoStop}%
	\bibitem [{\citenamefont {Scully}(1997)}]{Scully-quantum_optics}%
	\BibitemOpen
	\bibfield  {author} {\bibinfo {author} {\bibfnamefont {M.~O.}\ \bibnamefont
			{Scully}},\ }\href@noop {} {\emph {\bibinfo {title} {Quantum optics}}}\
	(\bibinfo  {publisher} {Cambridge University Press},\ \bibinfo {address}
	{Cambridge},\ \bibinfo {year} {1997})\BibitemShut {NoStop}%
	\bibitem [{\citenamefont {Lee}\ \emph {et~al.}(2014)\citenamefont {Lee},
		\citenamefont {Chan},\ and\ \citenamefont
		{Yelin}}]{LMG-collective_and_independent-decay-Lee}%
	\BibitemOpen
	\bibfield  {author} {\bibinfo {author} {\bibfnamefont {T.~E.}\ \bibnamefont
			{Lee}}, \bibinfo {author} {\bibfnamefont {C.-K.}\ \bibnamefont {Chan}}, \
		and\ \bibinfo {author} {\bibfnamefont {S.~F.}\ \bibnamefont {Yelin}},\ }\href
	{\doibase 10.1103/PhysRevA.90.052109} {\bibfield  {journal} {\bibinfo
			{journal} {Phys. Rev. A}\ }\textbf {\bibinfo {volume} {90}},\ \bibinfo
		{pages} {052109} (\bibinfo {year} {2014})}\BibitemShut {NoStop}%
	\bibitem [{\citenamefont {Kopylov}\ \emph {et~al.}(2013)\citenamefont
		{Kopylov}, \citenamefont {Emary},\ and\ \citenamefont
		{Brandes}}]{Kopylov_Counting-statistics-Dicke}%
	\BibitemOpen
	\bibfield  {author} {\bibinfo {author} {\bibfnamefont {W.}~\bibnamefont
			{Kopylov}}, \bibinfo {author} {\bibfnamefont {C.}~\bibnamefont {Emary}}, \
		and\ \bibinfo {author} {\bibfnamefont {T.}~\bibnamefont {Brandes}},\ }\href
	{\doibase 10.1103/PhysRevA.87.043840} {\bibfield  {journal} {\bibinfo
			{journal} {Phys. Rev. A}\ }\textbf {\bibinfo {volume} {87}},\ \bibinfo
		{pages} {043840} (\bibinfo {year} {2013})}\BibitemShut {NoStop}%
	\bibitem [{\citenamefont {Mostame}\ \emph {et~al.}(2007)\citenamefont
		{Mostame}, \citenamefont {Schaller},\ and\ \citenamefont
		{Sch\"utzhold}}]{mostame2007a}%
	\BibitemOpen
	\bibfield  {author} {\bibinfo {author} {\bibfnamefont {S.}~\bibnamefont
			{Mostame}}, \bibinfo {author} {\bibfnamefont {G.}~\bibnamefont {Schaller}}, \
		and\ \bibinfo {author} {\bibfnamefont {R.}~\bibnamefont {Sch\"utzhold}},\
	}\href@noop {} {\bibfield  {journal} {\bibinfo  {journal} {Physical Review
				A}\ }\textbf {\bibinfo {volume} {76}},\ \bibinfo {pages} {030304(R)}
		(\bibinfo {year} {2007})}\BibitemShut {NoStop}%
	\bibitem [{\citenamefont {Mostame}\ \emph {et~al.}(2010)\citenamefont
		{Mostame}, \citenamefont {Schaller},\ and\ \citenamefont
		{Sch\"utzhold}}]{mostame2010a}%
	\BibitemOpen
	\bibfield  {author} {\bibinfo {author} {\bibfnamefont {S.}~\bibnamefont
			{Mostame}}, \bibinfo {author} {\bibfnamefont {G.}~\bibnamefont {Schaller}}, \
		and\ \bibinfo {author} {\bibfnamefont {R.}~\bibnamefont {Sch\"utzhold}},\
	}\href@noop {} {\bibfield  {journal} {\bibinfo  {journal} {Physical Review
				A}\ }\textbf {\bibinfo {volume} {81}},\ \bibinfo {pages} {032305} (\bibinfo
		{year} {2010})}\BibitemShut {NoStop}%
	\bibitem [{\citenamefont {Klinder}\ \emph {et~al.}(2015)\citenamefont
		{Klinder}, \citenamefont {Ke{\ss}ler}, \citenamefont {Wolke}, \citenamefont
		{Mathey},\ and\ \citenamefont
		{Hemmerich}}]{Dicke_Dynamical_phase_transition_open_Dicke-Klinder}%
	\BibitemOpen
	\bibfield  {author} {\bibinfo {author} {\bibfnamefont {J.}~\bibnamefont
			{Klinder}}, \bibinfo {author} {\bibfnamefont {H.}~\bibnamefont {Ke{\ss}ler}},
		\bibinfo {author} {\bibfnamefont {M.}~\bibnamefont {Wolke}}, \bibinfo
		{author} {\bibfnamefont {L.}~\bibnamefont {Mathey}}, \ and\ \bibinfo {author}
		{\bibfnamefont {A.}~\bibnamefont {Hemmerich}},\ }\href {\doibase
		10.1073/pnas.1417132112} {\bibfield  {journal} {\bibinfo  {journal}
			{Proceedings of the National Academy of Sciences}\ }\textbf {\bibinfo
			{volume} {112}},\ \bibinfo {pages} {3290} (\bibinfo {year}
		{2015})}\BibitemShut {NoStop}%
	\bibitem [{\citenamefont {Lebreuilly}\ \emph {et~al.}()\citenamefont
		{Lebreuilly}, \citenamefont {Chiocchetta},\ and\ \citenamefont
		{Carusotto}}]{Thermalization-Pseudo_NonMarkovian_dissipative_system-Chiocchetta}%
	\BibitemOpen
	\bibfield  {author} {\bibinfo {author} {\bibfnamefont {J.}~\bibnamefont
			{Lebreuilly}}, \bibinfo {author} {\bibfnamefont {A.}~\bibnamefont
			{Chiocchetta}}, \ and\ \bibinfo {author} {\bibfnamefont {I.}~\bibnamefont
			{Carusotto}},\ }\href@noop {} {\bibinfo  {journal} {Phys. Rev. A}\
	}\BibitemShut {NoStop}%
	\bibitem [{\citenamefont {Faulstich}\ \emph {et~al.}(2017)\citenamefont
		{Faulstich}, \citenamefont {Kraft},\ and\ \citenamefont
		{Carmele}}]{Feedback-mirror_propiertes_as_time_delay_fb-carmele}%
	\BibitemOpen
	\bibfield  {journal} {  }\bibfield  {author} {\bibinfo {author} {\bibfnamefont
			{F.~M.}\ \bibnamefont {Faulstich}}, \bibinfo {author} {\bibfnamefont
			{M.}~\bibnamefont {Kraft}}, \ and\ \bibinfo {author} {\bibfnamefont
			{A.}~\bibnamefont {Carmele}},\ }\href@noop {} {\bibfield  {journal} {\bibinfo
			{journal} {Journal of Modern Optics}\ }\textbf {\bibinfo {volume} {65}},\
		\bibinfo {pages} {1323} (\bibinfo {year} {2017})}\BibitemShut {NoStop}%
	\bibitem [{\citenamefont {Kopylov}\ and\ \citenamefont
		{Brandes}(2015)}]{Kopylov-LMG_ESQPT_control}%
	\BibitemOpen
	\bibfield  {author} {\bibinfo {author} {\bibfnamefont {W.}~\bibnamefont
			{Kopylov}}\ and\ \bibinfo {author} {\bibfnamefont {T.}~\bibnamefont
			{Brandes}},\ }\href@noop {} {\bibfield  {journal} {\bibinfo  {journal} {New
				Journal of Physics}\ }\textbf {\bibinfo {volume} {17}},\ \bibinfo {pages}
		{103031} (\bibinfo {year} {2015})}\BibitemShut {NoStop}%
	\bibitem [{\citenamefont {Kabuss}\ \emph {et~al.}(2016)\citenamefont {Kabuss},
		\citenamefont {Katsch}, \citenamefont {Knorr},\ and\ \citenamefont
		{Carmele}}]{Pyragas-Uravaling_coherent_FB-Kabuss}%
	\BibitemOpen
	\bibfield  {author} {\bibinfo {author} {\bibfnamefont {J.}~\bibnamefont
			{Kabuss}}, \bibinfo {author} {\bibfnamefont {F.}~\bibnamefont {Katsch}},
		\bibinfo {author} {\bibfnamefont {A.}~\bibnamefont {Knorr}}, \ and\ \bibinfo
		{author} {\bibfnamefont {A.}~\bibnamefont {Carmele}},\ }\href {\doibase
		10.1364/JOSAB.33.000C10} {\bibfield  {journal} {\bibinfo  {journal} {J. Opt.
				Soc. Am. B}\ }\textbf {\bibinfo {volume} {33}},\ \bibinfo {pages} {C10}
		(\bibinfo {year} {2016})}\BibitemShut {NoStop}%
	\bibitem [{\citenamefont {Breuer}\ and\ \citenamefont
		{Petruccione}(2002)}]{Breuer-open_quantum_systems}%
	\BibitemOpen
	\bibfield  {author} {\bibinfo {author} {\bibfnamefont {H.-P.}\ \bibnamefont
			{Breuer}}\ and\ \bibinfo {author} {\bibfnamefont {F.}~\bibnamefont
			{Petruccione}},\ }\href@noop {} {\emph {\bibinfo {title} {The theory of open
				quantum systems}}}\ (\bibinfo  {publisher} {Oxford university press},\
	\bibinfo {year} {2002})\BibitemShut {NoStop}%
	\bibitem [{\citenamefont {Vogl}\ \emph {et~al.}(2012)\citenamefont {Vogl},
		\citenamefont {Schaller},\ and\ \citenamefont {Brandes}}]{vogl2012b}%
	\BibitemOpen
	\bibfield  {author} {\bibinfo {author} {\bibfnamefont {M.}~\bibnamefont
			{Vogl}}, \bibinfo {author} {\bibfnamefont {G.}~\bibnamefont {Schaller}}, \
		and\ \bibinfo {author} {\bibfnamefont {T.}~\bibnamefont {Brandes}},\ }\href
	{\doibase 10.1103/PhysRevLett.109.240402} {\bibfield  {journal} {\bibinfo
			{journal} {Physical Review Letters}\ }\textbf {\bibinfo {volume} {109}},\
		\bibinfo {pages} {240402} (\bibinfo {year} {2012})}\BibitemShut {NoStop}%
	\bibitem [{\citenamefont {Schaller}\ \emph {et~al.}(2014)\citenamefont
		{Schaller}, \citenamefont {Vogl},\ and\ \citenamefont
		{Brandes}}]{schaller2014a}%
	\BibitemOpen
	\bibfield  {author} {\bibinfo {author} {\bibfnamefont {G.}~\bibnamefont
			{Schaller}}, \bibinfo {author} {\bibfnamefont {M.}~\bibnamefont {Vogl}}, \
		and\ \bibinfo {author} {\bibfnamefont {T.}~\bibnamefont {Brandes}},\ }\href
	{http://stacks.iop.org/0953-8984/26/i=26/a=265001} {\bibfield  {journal}
		{\bibinfo  {journal} {Journal of Physics: Condensed Matter}\ }\textbf
		{\bibinfo {volume} {26}},\ \bibinfo {pages} {265001} (\bibinfo {year}
		{2014})}\BibitemShut {NoStop}%
	\bibitem [{\citenamefont {Schaller}(2014)}]{Schaller-QS_far_from_equilibrium}%
	\BibitemOpen
	\bibfield  {author} {\bibinfo {author} {\bibfnamefont {G.}~\bibnamefont
			{Schaller}},\ }\href@noop {} {\emph {\bibinfo {title} {Open Quantum Systems
				Far from Equilibrium}}}\ (\bibinfo  {publisher} {Springer},\ \bibinfo {year}
	{2014})\BibitemShut {NoStop}%
	\bibitem [{\citenamefont {Nazir}\ and\ \citenamefont
		{Schaller}(2019)}]{nazir2018a}%
	\BibitemOpen
	\bibfield  {author} {\bibinfo {author} {\bibfnamefont {A.}~\bibnamefont
			{Nazir}}\ and\ \bibinfo {author} {\bibfnamefont {G.}~\bibnamefont
			{Schaller}},\ }in\ \href {\doibase 10.1007/978-3-319-99046-0} {\emph
		{\bibinfo {booktitle} {Thermodynamics in the quantum regime }}},\ \bibinfo {series and number} {
		  \underline{ Recent progress and outlook}, Fundamental Theories
		of Physics},\ \bibinfo {editor} {edited by\ \bibinfo {editor} {\bibfnamefont
			{F.}~\bibnamefont {Binder}}, \bibinfo {editor} {\bibfnamefont {L.~A.}\
			\bibnamefont {Correa}}, \bibinfo {editor} {\bibfnamefont {C.}~\bibnamefont
			{Gogolin}}, \bibinfo {editor} {\bibfnamefont {J.}~\bibnamefont {Anders}}, \
		and\ \bibinfo {editor} {\bibfnamefont {G.}~\bibnamefont {Adesso}}}\ (\bibinfo
	{publisher} {Springer},\ \bibinfo {year} {2019})\BibitemShut {NoStop}%
	\bibitem [{\citenamefont {Strasberg}\ \emph {et~al.}(2016)\citenamefont
		{Strasberg}, \citenamefont {Schaller}, \citenamefont {Lambert},\ and\
		\citenamefont
		{Brandes}}]{Thermodynamics-Nonequilibr_react_coordinate-Strasberg}%
	\BibitemOpen
	\bibfield  {author} {\bibinfo {author} {\bibfnamefont {P.}~\bibnamefont
			{Strasberg}}, \bibinfo {author} {\bibfnamefont {G.}~\bibnamefont {Schaller}},
		\bibinfo {author} {\bibfnamefont {N.}~\bibnamefont {Lambert}}, \ and\
		\bibinfo {author} {\bibfnamefont {T.}~\bibnamefont {Brandes}},\ }\href
	{\doibase 10.1088/1367-2630/18/7/073007} {\bibfield  {journal} {\bibinfo
			{journal} {New Journal of Physics}\ }\textbf {\bibinfo {volume} {18}},\
		\bibinfo {pages} {073007} (\bibinfo {year} {2016})}\BibitemShut {NoStop}%
	\bibitem [{\citenamefont {Meshkov}\ and\ \citenamefont
		{Glick}(1965)}]{LMG-validity_many_body_approx-Lipkin}%
	\BibitemOpen
	\bibfield  {author} {\bibinfo {author} {\bibfnamefont {N.}~\bibnamefont
			{Meshkov}}\ and\ \bibinfo {author} {\bibfnamefont {A.}~\bibnamefont
			{Glick}},\ }\href@noop {} {\bibfield  {journal} {\bibinfo  {journal} {Nuclear
				physics.}\ }\textbf {\bibinfo {volume} {62}} (\bibinfo {year}
		{1965})}\BibitemShut {NoStop}%
	\bibitem [{\citenamefont {Mahan}(2013)}]{mahan2013many}%
	\BibitemOpen
	\bibfield  {author} {\bibinfo {author} {\bibfnamefont {G.~D.}\ \bibnamefont
			{Mahan}},\ }\href@noop {} {\emph {\bibinfo {title} {Many-particle physics}}}\
	(\bibinfo  {publisher} {Springer Science \& Business Media},\ \bibinfo {year}
	{2013})\BibitemShut {NoStop}%
	\bibitem [{\citenamefont {Glazman}\ and\ \citenamefont
		{Shekhter}(1988)}]{Polaron-inelastic_resonant_tunneling_electrons_barrier-Glazman}%
	\BibitemOpen
	\bibfield  {author} {\bibinfo {author} {\bibfnamefont {L.}~\bibnamefont
			{Glazman}}\ and\ \bibinfo {author} {\bibfnamefont {R.}~\bibnamefont
			{Shekhter}},\ }\href@noop {} {\bibfield  {journal} {\bibinfo  {journal} {Sov.
				Phys. JETP}\ }\textbf {\bibinfo {volume} {67}},\ \bibinfo {pages} {163}
		(\bibinfo {year} {1988})}\BibitemShut {NoStop}%
	\bibitem [{\citenamefont {Wingreen}\ \emph {et~al.}(1988)\citenamefont
		{Wingreen}, \citenamefont {Jacobsen},\ and\ \citenamefont
		{Wilkins}}]{Polaron-electron_phonon_resonant_tunneling-wingreen}%
	\BibitemOpen
	\bibfield  {author} {\bibinfo {author} {\bibfnamefont {N.~S.}\ \bibnamefont
			{Wingreen}}, \bibinfo {author} {\bibfnamefont {K.~W.}\ \bibnamefont
			{Jacobsen}}, \ and\ \bibinfo {author} {\bibfnamefont {J.~W.}\ \bibnamefont
			{Wilkins}},\ }\href {\doibase 10.1103/PhysRevLett.61.1396} {\bibfield
		{journal} {\bibinfo  {journal} {Phys. Rev. Lett.}\ }\textbf {\bibinfo
			{volume} {61}},\ \bibinfo {pages} {1396} (\bibinfo {year}
		{1988})}\BibitemShut {NoStop}%
	\bibitem [{\citenamefont
		{Brandes}(2005)}]{Polaron-coherent_collective_effects-_mesoscopic-Brandes}%
	\BibitemOpen
	\bibfield  {author} {\bibinfo {author} {\bibfnamefont {T.}~\bibnamefont
			{Brandes}},\ }\href {\doibase 10.1016/j.physrep.2004.12.002} {\bibfield
		{journal} {\bibinfo  {journal} {Physics Reports}\ }\textbf {\bibinfo {volume}
			{408}},\ \bibinfo {pages} {315} (\bibinfo {year} {2005})}\BibitemShut
	{NoStop}%
	\bibitem [{\citenamefont {Schaller}\ \emph {et~al.}(2013)\citenamefont
		{Schaller}, \citenamefont {Krause}, \citenamefont {Brandes},\ and\
		\citenamefont
		{Esposito}}]{Polaron-electron_transistor_strong_coupling_counting-Schaller}%
	\BibitemOpen
	\bibfield  {author} {\bibinfo {author} {\bibfnamefont {G.}~\bibnamefont
			{Schaller}}, \bibinfo {author} {\bibfnamefont {T.}~\bibnamefont {Krause}},
		\bibinfo {author} {\bibfnamefont {T.}~\bibnamefont {Brandes}}, \ and\
		\bibinfo {author} {\bibfnamefont {M.}~\bibnamefont {Esposito}},\ }\href
	{\doibase 10.1088/1367-2630/15/3/033032} {\bibfield  {journal} {\bibinfo
			{journal} {New Journal of Physics}\ }\textbf {\bibinfo {volume} {15}},\
		\bibinfo {pages} {033032} (\bibinfo {year} {2013})}\BibitemShut {NoStop}%
	\bibitem [{\citenamefont {Thorwart}\ \emph {et~al.}(2004)\citenamefont
		{Thorwart}, \citenamefont {Paladino},\ and\ \citenamefont
		{Grifoni}}]{Polaro-spin_boson_dynamics-Thorwart}%
	\BibitemOpen
	\bibfield  {author} {\bibinfo {author} {\bibfnamefont {M.}~\bibnamefont
			{Thorwart}}, \bibinfo {author} {\bibfnamefont {E.}~\bibnamefont {Paladino}},
		\ and\ \bibinfo {author} {\bibfnamefont {M.}~\bibnamefont {Grifoni}},\
	}\href@noop {} {\bibfield  {journal} {\bibinfo  {journal} {Chemical Physics}\
		}\textbf {\bibinfo {volume} {296}},\ \bibinfo {pages} {333} (\bibinfo {year}
		{2004})}\BibitemShut {NoStop}%
	\bibitem [{\citenamefont {Wilhelm}\ \emph {et~al.}(2004)\citenamefont
		{Wilhelm}, \citenamefont {Kleff},\ and\ \citenamefont
		{Von~Delft}}]{Polaron-spin_boson_comparisson-Wilhelm}%
	\BibitemOpen
	\bibfield  {author} {\bibinfo {author} {\bibfnamefont {F.}~\bibnamefont
			{Wilhelm}}, \bibinfo {author} {\bibfnamefont {S.}~\bibnamefont {Kleff}}, \
		and\ \bibinfo {author} {\bibfnamefont {J.}~\bibnamefont {Von~Delft}},\
	}\href@noop {} {\bibfield  {journal} {\bibinfo  {journal} {Chemical physics}\
		}\textbf {\bibinfo {volume} {296}},\ \bibinfo {pages} {345} (\bibinfo {year}
		{2004})}\BibitemShut {NoStop}%
	\bibitem [{\citenamefont {Brandes}\ and\ \citenamefont
		{Vorrath}(2003)}]{Polaron-dissipation_two_level_comparisson-brandes}%
	\BibitemOpen
	\bibfield  {author} {\bibinfo {author} {\bibfnamefont {T.}~\bibnamefont
			{Brandes}}\ and\ \bibinfo {author} {\bibfnamefont {T.}~\bibnamefont
			{Vorrath}},\ }\href@noop {} {\bibfield  {journal} {\bibinfo  {journal}
			{International Journal of Modern Physics B}\ }\textbf {\bibinfo {volume}
			{17}},\ \bibinfo {pages} {5465} (\bibinfo {year} {2003})}\BibitemShut
	{NoStop}%
	\bibitem [{\citenamefont {Alcalde}\ \emph {et~al.}(2012)\citenamefont
		{Alcalde}, \citenamefont {Bucher}, \citenamefont {Emary},\ and\ \citenamefont
		{Brandes}}]{Dicke_ultra-strong-coupling-limit-qpt_Bucher}%
	\BibitemOpen
	\bibfield  {author} {\bibinfo {author} {\bibfnamefont {M.~A.}\ \bibnamefont
			{Alcalde}}, \bibinfo {author} {\bibfnamefont {M.}~\bibnamefont {Bucher}},
		\bibinfo {author} {\bibfnamefont {C.}~\bibnamefont {Emary}}, \ and\ \bibinfo
		{author} {\bibfnamefont {T.}~\bibnamefont {Brandes}},\ }\href {\doibase
		10.1103/PhysRevE.86.012101} {\bibfield  {journal} {\bibinfo  {journal} {Phys.
				Rev. E}\ }\textbf {\bibinfo {volume} {86}},\ \bibinfo {pages} {012101}
		(\bibinfo {year} {2012})}\BibitemShut {NoStop}%
	\bibitem [{\citenamefont {Krause}\ \emph {et~al.}(2015)\citenamefont {Krause},
		\citenamefont {Brandes}, \citenamefont {Esposito},\ and\ \citenamefont
		{Schaller}}]{Krause2015}%
	\BibitemOpen
	\bibfield  {author} {\bibinfo {author} {\bibfnamefont {T.}~\bibnamefont
			{Krause}}, \bibinfo {author} {\bibfnamefont {T.}~\bibnamefont {Brandes}},
		\bibinfo {author} {\bibfnamefont {M.}~\bibnamefont {Esposito}}, \ and\
		\bibinfo {author} {\bibfnamefont {G.}~\bibnamefont {Schaller}},\ }\href
	{\doibase 10.1063/1.4916359} {\bibfield  {journal} {\bibinfo  {journal} {The
				Journal of Chemical Physics}\ }\textbf {\bibinfo {volume} {142}},\ \bibinfo
		{pages} {134106} (\bibinfo {year} {2015})}\BibitemShut {NoStop}%
	\bibitem [{\citenamefont {Kirton}\ and\ \citenamefont
		{Keeling}(2013)}]{Keeling_PRL-nonequilibrium_model_photon-cond}%
	\BibitemOpen
	\bibfield  {author} {\bibinfo {author} {\bibfnamefont {P.}~\bibnamefont
			{Kirton}}\ and\ \bibinfo {author} {\bibfnamefont {J.}~\bibnamefont
			{Keeling}},\ }\href {\doibase 10.1103/PhysRevLett.111.100404} {\bibfield
		{journal} {\bibinfo  {journal} {Phys. Rev. Lett.}\ }\textbf {\bibinfo
			{volume} {111}},\ \bibinfo {pages} {100404} (\bibinfo {year}
		{2013})}\BibitemShut {NoStop}%
	\bibitem [{\citenamefont {Radonji{\'{c}}}\ \emph {et~al.}(2018)\citenamefont
		{Radonji{\'{c}}}, \citenamefont {Kopylov}, \citenamefont {Bala{\v{z}}},\ and\
		\citenamefont {Pelster}}]{PhotCond-interplay_coh_disisp_dynamics-Milan}%
	\BibitemOpen
	\bibfield  {author} {\bibinfo {author} {\bibfnamefont {M.}~\bibnamefont
			{Radonji{\'{c}}}}, \bibinfo {author} {\bibfnamefont {W.}~\bibnamefont
			{Kopylov}}, \bibinfo {author} {\bibfnamefont {A.}~\bibnamefont
			{Bala{\v{z}}}}, \ and\ \bibinfo {author} {\bibfnamefont {A.}~\bibnamefont
			{Pelster}},\ }\href {\doibase 10.1088/1367-2630/aac2a6} {\bibfield  {journal}
		{\bibinfo  {journal} {New Journal of Physics}\ }\textbf {\bibinfo {volume}
			{20}},\ \bibinfo {pages} {055014} (\bibinfo {year} {2018})}\BibitemShut
	{NoStop}%
	\bibitem [{\citenamefont {Bhaseen}\ \emph {et~al.}(2012)\citenamefont
		{Bhaseen}, \citenamefont {Mayoh}, \citenamefont {Simons},\ and\ \citenamefont
		{Keeling}}]{Bhaseen_dynamics_of_nonequilibrium_dicke_models}%
	\BibitemOpen
	\bibfield  {author} {\bibinfo {author} {\bibfnamefont {M.~J.}\ \bibnamefont
			{Bhaseen}}, \bibinfo {author} {\bibfnamefont {J.}~\bibnamefont {Mayoh}},
		\bibinfo {author} {\bibfnamefont {B.~D.}\ \bibnamefont {Simons}}, \ and\
		\bibinfo {author} {\bibfnamefont {J.}~\bibnamefont {Keeling}},\ }\href
	{\doibase 10.1103/PhysRevA.85.013817} {\bibfield  {journal} {\bibinfo
			{journal} {Phys. Rev. A}\ }\textbf {\bibinfo {volume} {85}},\ \bibinfo
		{pages} {013817} (\bibinfo {year} {2012})}\BibitemShut {NoStop}%
	\bibitem [{\citenamefont {Nagy}\ \emph {et~al.}(2011)\citenamefont {Nagy},
		\citenamefont {Szirmai},\ and\ \citenamefont
		{Domokos}}]{Dicke_open-critical_exponent_of_noise-Nagy}%
	\BibitemOpen
	\bibfield  {author} {\bibinfo {author} {\bibfnamefont {D.}~\bibnamefont
			{Nagy}}, \bibinfo {author} {\bibfnamefont {G.}~\bibnamefont {Szirmai}}, \
		and\ \bibinfo {author} {\bibfnamefont {P.}~\bibnamefont {Domokos}},\ }\href
	{\doibase 10.1103/PhysRevA.84.043637} {\bibfield  {journal} {\bibinfo
			{journal} {Phys. Rev. A}\ }\textbf {\bibinfo {volume} {84}},\ \bibinfo
		{pages} {043637} (\bibinfo {year} {2011})}\BibitemShut {NoStop}%
	\bibitem [{\citenamefont {Hwang}\ \emph {et~al.}(2018)\citenamefont {Hwang},
		\citenamefont {Rabl},\ and\ \citenamefont
		{Plenio}}]{Rabi_dissipative-QPT_Plenio}%
	\BibitemOpen
	\bibfield  {author} {\bibinfo {author} {\bibfnamefont {M.-J.}\ \bibnamefont
			{Hwang}}, \bibinfo {author} {\bibfnamefont {P.}~\bibnamefont {Rabl}}, \ and\
		\bibinfo {author} {\bibfnamefont {M.~B.}\ \bibnamefont {Plenio}},\ }\href
	{\doibase 10.1103/PhysRevA.97.013825} {\bibfield  {journal} {\bibinfo
			{journal} {Phys. Rev. A}\ }\textbf {\bibinfo {volume} {97}},\ \bibinfo
		{pages} {013825} (\bibinfo {year} {2018})}\BibitemShut {NoStop}%
	\bibitem [{\citenamefont {Gelhausen}\ and\ \citenamefont
		{Buchhold}(2018)}]{Dicke-dissipative_bistability_noise_nonthermal-Buchhold}%
	\BibitemOpen
	\bibfield  {author} {\bibinfo {author} {\bibfnamefont {J.}~\bibnamefont
			{Gelhausen}}\ and\ \bibinfo {author} {\bibfnamefont {M.}~\bibnamefont
			{Buchhold}},\ }\href {\doibase 10.1103/PhysRevA.97.023807} {\bibfield
		{journal} {\bibinfo  {journal} {Phys. Rev. A}\ }\textbf {\bibinfo {volume}
			{97}},\ \bibinfo {pages} {023807} (\bibinfo {year} {2018})}\BibitemShut
	{NoStop}%
	\bibitem [{\citenamefont {Li}\ \emph {et~al.}(2017)\citenamefont {Li},
		\citenamefont {Piryatinski}, \citenamefont {Jerke}, \citenamefont {Kandada},
		\citenamefont {Silva},\ and\ \citenamefont
		{Bittner}}]{Dicke-dynamical_symmetry_breaking-Li}%
	\BibitemOpen
	\bibfield  {author} {\bibinfo {author} {\bibfnamefont {H.}~\bibnamefont
			{Li}}, \bibinfo {author} {\bibfnamefont {A.}~\bibnamefont {Piryatinski}},
		\bibinfo {author} {\bibfnamefont {J.}~\bibnamefont {Jerke}}, \bibinfo
		{author} {\bibfnamefont {A.~R.~S.}\ \bibnamefont {Kandada}}, \bibinfo
		{author} {\bibfnamefont {C.}~\bibnamefont {Silva}}, \ and\ \bibinfo {author}
		{\bibfnamefont {E.~R.}\ \bibnamefont {Bittner}},\ }\href {\doibase
		10.1088/2058-9565/aa93b6} {\bibfield  {journal} {\bibinfo  {journal} {Quantum
				Science and Technology}\ }\textbf {\bibinfo {volume} {3}},\ \bibinfo {pages}
		{015003} (\bibinfo {year} {2017})}\BibitemShut {NoStop}%
	\bibitem [{\citenamefont {Morrison}\ and\ \citenamefont
		{Parkins}(2008)}]{Morrison-Dissipative_LMG-and_QPT}%
	\BibitemOpen
	\bibfield  {author} {\bibinfo {author} {\bibfnamefont {S.}~\bibnamefont
			{Morrison}}\ and\ \bibinfo {author} {\bibfnamefont {A.~S.}\ \bibnamefont
			{Parkins}},\ }\href {\doibase 10.1103/PhysRevLett.100.040403} {\bibfield
		{journal} {\bibinfo  {journal} {Phys. Rev. Lett.}\ }\textbf {\bibinfo
			{volume} {100}},\ \bibinfo {pages} {040403} (\bibinfo {year}
		{2008})}\BibitemShut {NoStop}%
	\bibitem [{\citenamefont {Or\'us}\ \emph {et~al.}(2008)\citenamefont {Or\'us},
		\citenamefont {Dusuel},\ and\ \citenamefont
		{Vidal}}]{LMG-Critical_scaling_law_entaglement-Vidal}%
	\BibitemOpen
	\bibfield  {author} {\bibinfo {author} {\bibfnamefont {R.}~\bibnamefont
			{Or\'us}}, \bibinfo {author} {\bibfnamefont {S.}~\bibnamefont {Dusuel}}, \
		and\ \bibinfo {author} {\bibfnamefont {J.}~\bibnamefont {Vidal}},\ }\href
	{\doibase 10.1103/PhysRevLett.101.025701} {\bibfield  {journal} {\bibinfo
			{journal} {Phys. Rev. Lett.}\ }\textbf {\bibinfo {volume} {101}},\ \bibinfo
		{pages} {025701} (\bibinfo {year} {2008})}\BibitemShut {NoStop}%
	\bibitem [{\citenamefont {Kochma\'nski}\ \emph {et~al.}(2013)\citenamefont
		{Kochma\'nski}, \citenamefont {Paszkiewicz},\ and\ \citenamefont
		{Wolski}}]{kochmanski2013a}%
	\BibitemOpen
	\bibfield  {author} {\bibinfo {author} {\bibfnamefont {M.}~\bibnamefont
			{Kochma\'nski}}, \bibinfo {author} {\bibfnamefont {T.}~\bibnamefont
			{Paszkiewicz}}, \ and\ \bibinfo {author} {\bibfnamefont {S.}~\bibnamefont
			{Wolski}},\ }\href {\doibase 10.1088/0143-0807/34/6/1555} {\bibfield
		{journal} {\bibinfo  {journal} {European Journal of Physics}\ }\textbf
		{\bibinfo {volume} {34}},\ \bibinfo {pages} {1555} (\bibinfo {year}
		{2013})}\BibitemShut {NoStop}%
	\bibitem [{\citenamefont {Lipkin}\ \emph {et~al.}(1965)\citenamefont {Lipkin},
		\citenamefont {Meshkov},\ and\ \citenamefont
		{Glick}}]{LMG-lipkin1965validity}%
	\BibitemOpen
	\bibfield  {author} {\bibinfo {author} {\bibfnamefont {H.~J.}\ \bibnamefont
			{Lipkin}}, \bibinfo {author} {\bibfnamefont {N.}~\bibnamefont {Meshkov}}, \
		and\ \bibinfo {author} {\bibfnamefont {A.}~\bibnamefont {Glick}},\
	}\href@noop {} {\bibfield  {journal} {\bibinfo  {journal} {Nucl. Phys.}\
		}\textbf {\bibinfo {volume} {62}},\ \bibinfo {pages} {188} (\bibinfo {year}
		{1965})}\BibitemShut {NoStop}%
	\bibitem [{\citenamefont {Glick}\ \emph {et~al.}(1965)\citenamefont {Glick},
		\citenamefont {Lipkin},\ and\ \citenamefont
		{Meshkov}}]{LMG-validity_many_body_approx-Lipkin-3}%
	\BibitemOpen
	\bibfield  {author} {\bibinfo {author} {\bibfnamefont {A.}~\bibnamefont
			{Glick}}, \bibinfo {author} {\bibfnamefont {H.}~\bibnamefont {Lipkin}}, \
		and\ \bibinfo {author} {\bibfnamefont {N.}~\bibnamefont {Meshkov}},\ }\href
	{\doibase http://dx.doi.org/10.1016/0029-5582(65)90864-3} {\bibfield
		{journal} {\bibinfo  {journal} {Nuclear Physics}\ }\textbf {\bibinfo {volume}
			{62}},\ \bibinfo {pages} {211 } (\bibinfo {year} {1965})}\BibitemShut
	{NoStop}%
	\bibitem [{\citenamefont {Gilmore}\ and\ \citenamefont
		{Feng}(1978)}]{LMG-phase_transition-Gilmore}%
	\BibitemOpen
	\bibfield  {author} {\bibinfo {author} {\bibfnamefont {R.}~\bibnamefont
			{Gilmore}}\ and\ \bibinfo {author} {\bibfnamefont {D.}~\bibnamefont {Feng}},\
	}\href@noop {} {\bibfield  {journal} {\bibinfo  {journal} {Nuclear Physics
				A}\ }\textbf {\bibinfo {volume} {301}},\ \bibinfo {pages} {189} (\bibinfo
		{year} {1978})}\BibitemShut {NoStop}%
	\bibitem [{\citenamefont {Leyvraz}\ and\ \citenamefont
		{Heiss}(2005)}]{LMG-large_N_scaling_behaviour-Heiss}%
	\BibitemOpen
	\bibfield  {author} {\bibinfo {author} {\bibfnamefont {F.}~\bibnamefont
			{Leyvraz}}\ and\ \bibinfo {author} {\bibfnamefont {W.~D.}\ \bibnamefont
			{Heiss}},\ }\href {\doibase 10.1103/PhysRevLett.95.050402} {\bibfield
		{journal} {\bibinfo  {journal} {Phys. Rev. Lett.}\ }\textbf {\bibinfo
			{volume} {95}},\ \bibinfo {pages} {050402} (\bibinfo {year}
		{2005})}\BibitemShut {NoStop}%
	\bibitem [{\citenamefont {Sorokin}\ \emph {et~al.}(2014)\citenamefont
		{Sorokin}, \citenamefont {Bastidas},\ and\ \citenamefont
		{Brandes}}]{LMG-networks_qpt-sorokin}%
	\BibitemOpen
	\bibfield  {author} {\bibinfo {author} {\bibfnamefont {A.~V.}\ \bibnamefont
			{Sorokin}}, \bibinfo {author} {\bibfnamefont {V.~M.}\ \bibnamefont
			{Bastidas}}, \ and\ \bibinfo {author} {\bibfnamefont {T.}~\bibnamefont
			{Brandes}},\ }\href {\doibase 10.1103/PhysRevE.90.042141} {\bibfield
		{journal} {\bibinfo  {journal} {Phys. Rev. E}\ }\textbf {\bibinfo {volume}
			{90}},\ \bibinfo {pages} {042141} (\bibinfo {year} {2014})}\BibitemShut
	{NoStop}%
	\bibitem [{\citenamefont {Vidal}\ \emph {et~al.}(2004)\citenamefont {Vidal},
		\citenamefont {Palacios},\ and\ \citenamefont
		{Aslangul}}]{LMG_Entanglement_dynamics_Vidal}%
	\BibitemOpen
	\bibfield  {author} {\bibinfo {author} {\bibfnamefont {J.}~\bibnamefont
			{Vidal}}, \bibinfo {author} {\bibfnamefont {G.}~\bibnamefont {Palacios}}, \
		and\ \bibinfo {author} {\bibfnamefont {C.}~\bibnamefont {Aslangul}},\ }\href
	{\doibase 10.1103/PhysRevA.70.062304} {\bibfield  {journal} {\bibinfo
			{journal} {Phys. Rev. A}\ }\textbf {\bibinfo {volume} {70}},\ \bibinfo
		{pages} {062304} (\bibinfo {year} {2004})}\BibitemShut {NoStop}%
	\bibitem [{\citenamefont {Huang}\ \emph {et~al.}(2018)\citenamefont {Huang},
		\citenamefont {Li},\ and\ \citenamefont
		{qi~Yin}}]{LMG-symmetry_breaking_dynamics_finite_size-Huang}%
	\BibitemOpen
	\bibfield  {author} {\bibinfo {author} {\bibfnamefont {Y.}~\bibnamefont
			{Huang}}, \bibinfo {author} {\bibfnamefont {T.}~\bibnamefont {Li}}, \ and\
		\bibinfo {author} {\bibfnamefont {Z.}~\bibnamefont {qi~Yin}},\ }\href
	{\doibase 10.1103/physreva.97.012115} {\bibfield  {journal} {\bibinfo
			{journal} {Physical Review A}\ }\textbf {\bibinfo {volume} {97}} (\bibinfo
		{year} {2018}),\ 10.1103/physreva.97.012115}\BibitemShut {NoStop}%
	\bibitem [{\citenamefont {Ribeiro}\ \emph {et~al.}(2008)\citenamefont
		{Ribeiro}, \citenamefont {Vidal},\ and\ \citenamefont
		{Mosseri}}]{LMG-spectrum_thermodynamic_limit_and_finite_size-corr-Mosseri}%
	\BibitemOpen
	\bibfield  {author} {\bibinfo {author} {\bibfnamefont {P.}~\bibnamefont
			{Ribeiro}}, \bibinfo {author} {\bibfnamefont {J.}~\bibnamefont {Vidal}}, \
		and\ \bibinfo {author} {\bibfnamefont {R.}~\bibnamefont {Mosseri}},\ }\href
	{\doibase 10.1103/PhysRevE.78.021106} {\bibfield  {journal} {\bibinfo
			{journal} {Phys. Rev. E}\ }\textbf {\bibinfo {volume} {78}},\ \bibinfo
		{pages} {021106} (\bibinfo {year} {2008})}\BibitemShut {NoStop}%
	\bibitem [{\citenamefont {Dusuel}\ and\ \citenamefont
		{Vidal}(2005)}]{LMG-Finite_size_scalling_Dusuel}%
	\BibitemOpen
	\bibfield  {author} {\bibinfo {author} {\bibfnamefont {S.}~\bibnamefont
			{Dusuel}}\ and\ \bibinfo {author} {\bibfnamefont {J.}~\bibnamefont {Vidal}},\
	}\href {\doibase 10.1103/PhysRevB.71.224420} {\bibfield  {journal} {\bibinfo
			{journal} {Phys. Rev. B}\ }\textbf {\bibinfo {volume} {71}},\ \bibinfo
		{pages} {224420} (\bibinfo {year} {2005})}\BibitemShut {NoStop}%
	\bibitem [{\citenamefont {Dusuel}\ and\ \citenamefont
		{Vidal}(2004)}]{LMG-Finite_size_scaling-Vidal}%
	\BibitemOpen
	\bibfield  {author} {\bibinfo {author} {\bibfnamefont {S.}~\bibnamefont
			{Dusuel}}\ and\ \bibinfo {author} {\bibfnamefont {J.}~\bibnamefont {Vidal}},\
	}\href {\doibase 10.1103/PhysRevLett.93.237204} {\bibfield  {journal}
		{\bibinfo  {journal} {Phys. Rev. Lett.}\ }\textbf {\bibinfo {volume} {93}},\
		\bibinfo {pages} {237204} (\bibinfo {year} {2004})}\BibitemShut {NoStop}%
	\bibitem [{\citenamefont {Zimmermann}\ \emph {et~al.}(2018)\citenamefont
		{Zimmermann}, \citenamefont {Kopylov},\ and\ \citenamefont
		{Schaller}}]{LMG-Wiseman_control-Kopylov}%
	\BibitemOpen
	\bibfield  {author} {\bibinfo {author} {\bibfnamefont {S.}~\bibnamefont
			{Zimmermann}}, \bibinfo {author} {\bibfnamefont {W.}~\bibnamefont {Kopylov}},
		\ and\ \bibinfo {author} {\bibfnamefont {G.}~\bibnamefont {Schaller}},\
	}\href {\doibase 10.1088/1751-8121/aad2c3} {\bibfield  {journal} {\bibinfo
			{journal} {Journal of Physics A: Mathematical and Theoretical}\ }\textbf
		{\bibinfo {volume} {51}},\ \bibinfo {pages} {385301} (\bibinfo {year}
		{2018})}\BibitemShut {NoStop}%
	\bibitem [{\citenamefont {Holstein}\ and\ \citenamefont
		{Primakoff}(1940)}]{HP-Trafo_field_dependency_of_ferromagnet_Primakoff}%
	\BibitemOpen
	\bibfield  {author} {\bibinfo {author} {\bibfnamefont {T.}~\bibnamefont
			{Holstein}}\ and\ \bibinfo {author} {\bibfnamefont {H.}~\bibnamefont
			{Primakoff}},\ }\href {\doibase 10.1103/PhysRev.58.1098} {\bibfield
		{journal} {\bibinfo  {journal} {Phys. Rev.}\ }\textbf {\bibinfo {volume}
			{58}},\ \bibinfo {pages} {1098} (\bibinfo {year} {1940})}\BibitemShut
	{NoStop}%
	\bibitem [{\citenamefont {Emary}\ and\ \citenamefont
		{Brandes}(2003)}]{Clive-Brandes_Chaos_and_qpt_Dicke}%
	\BibitemOpen
	\bibfield  {author} {\bibinfo {author} {\bibfnamefont {C.}~\bibnamefont
			{Emary}}\ and\ \bibinfo {author} {\bibfnamefont {T.}~\bibnamefont
			{Brandes}},\ }\href {\doibase 10.1103/PhysRevE.67.066203} {\bibfield
		{journal} {\bibinfo  {journal} {Phys. Rev. E}\ }\textbf {\bibinfo {volume}
			{67}},\ \bibinfo {pages} {066203} (\bibinfo {year} {2003})}\BibitemShut
	{NoStop}%
	\bibitem [{\citenamefont {Louw}\ \emph {et~al.}(2019)\citenamefont {Louw},
		\citenamefont {Kriel},\ and\ \citenamefont
		{Kastner}}]{LMG_thermalization_kastner}%
	\BibitemOpen
	\bibfield  {author} {\bibinfo {author} {\bibfnamefont {J.~C.}\ \bibnamefont
			{Louw}}, \bibinfo {author} {\bibfnamefont {J.~N.}\ \bibnamefont {Kriel}}, \
		and\ \bibinfo {author} {\bibfnamefont {M.}~\bibnamefont {Kastner}},\ }\href
	{\doibase 10.1103/PhysRevA.100.022115} {\bibfield  {journal} {\bibinfo
			{journal} {Phys. Rev. A}\ }\textbf {\bibinfo {volume} {100}},\ \bibinfo
		{pages} {022115} (\bibinfo {year} {2019})}\BibitemShut {NoStop}%
	\bibitem [{\citenamefont {Dimer}\ \emph {et~al.}(2007)\citenamefont {Dimer},
		\citenamefont {Estienne}, \citenamefont {Parkins},\ and\ \citenamefont
		{Carmichael}}]{Dicke-realization_Dicke_in_cavity_system-Dimer}%
	\BibitemOpen
	\bibfield  {author} {\bibinfo {author} {\bibfnamefont {F.}~\bibnamefont
			{Dimer}}, \bibinfo {author} {\bibfnamefont {B.}~\bibnamefont {Estienne}},
		\bibinfo {author} {\bibfnamefont {A.~S.}\ \bibnamefont {Parkins}}, \ and\
		\bibinfo {author} {\bibfnamefont {H.~J.}\ \bibnamefont {Carmichael}},\ }\href
	{\doibase 10.1103/PhysRevA.75.013804} {\bibfield  {journal} {\bibinfo
			{journal} {Phys. Rev. A}\ }\textbf {\bibinfo {volume} {75}},\ \bibinfo
		{pages} {013804} (\bibinfo {year} {2007})}\BibitemShut {NoStop}%
	\bibitem [{\citenamefont {Wang}\ \emph {et~al.}(2015)\citenamefont {Wang},
		\citenamefont {Ren},\ and\ \citenamefont {Cao}}]{Polaron-spin_boson-Wang}%
	\BibitemOpen
	\bibfield  {author} {\bibinfo {author} {\bibfnamefont {C.}~\bibnamefont
			{Wang}}, \bibinfo {author} {\bibfnamefont {J.}~\bibnamefont {Ren}}, \ and\
		\bibinfo {author} {\bibfnamefont {J.}~\bibnamefont {Cao}},\ }\href {\doibase
		10.1038/srep11787} {\bibfield  {journal} {\bibinfo  {journal} {Scientific
				Reports}\ }\textbf {\bibinfo {volume} {5}} (\bibinfo {year} {2015}),\
		10.1038/srep11787}\BibitemShut {NoStop}%
	\bibitem [{\citenamefont {Wang}\ and\ \citenamefont
		{Sun}(2015)}]{Polaron_collective_system_with_interaction-Wang}%
	\BibitemOpen
	\bibfield  {author} {\bibinfo {author} {\bibfnamefont {C.}~\bibnamefont
			{Wang}}\ and\ \bibinfo {author} {\bibfnamefont {K.-W.}\ \bibnamefont {Sun}},\
	}\href {\doibase 10.1016/j.aop.2015.09.005} {\bibfield  {journal} {\bibinfo
			{journal} {Annals of Physics}\ }\textbf {\bibinfo {volume} {362}},\ \bibinfo
		{pages} {703} (\bibinfo {year} {2015})}\BibitemShut {NoStop}%
	\bibitem [{\citenamefont {Garg}\ \emph {et~al.}(1985)\citenamefont {Garg},
		\citenamefont {Onuchic},\ and\ \citenamefont
		{Ambegaokar}}]{Reaction_Coordinate-Effect_friction_electron_transfer_biomol-Garg}%
	\BibitemOpen
	\bibfield  {author} {\bibinfo {author} {\bibfnamefont {A.}~\bibnamefont
			{Garg}}, \bibinfo {author} {\bibfnamefont {J.~N.}\ \bibnamefont {Onuchic}}, \
		and\ \bibinfo {author} {\bibfnamefont {V.}~\bibnamefont {Ambegaokar}},\
	}\href {\doibase 10.1063/1.449017} {\bibfield  {journal} {\bibinfo  {journal}
			{The Journal of Chemical Physics}\ }\textbf {\bibinfo {volume} {83}},\
		\bibinfo {pages} {4491} (\bibinfo {year} {1985})}\BibitemShut {NoStop}%
	\bibitem [{\citenamefont
		{Tsallis}(1978)}]{Hamilt_diagonalization_quadratik_Tsallis1978}%
	\BibitemOpen
	\bibfield  {author} {\bibinfo {author} {\bibfnamefont {C.}~\bibnamefont
			{Tsallis}},\ }\href {\doibase 10.1063/1.523549} {\bibfield  {journal}
		{\bibinfo  {journal} {Journal of Mathematical Physics}\ }\textbf {\bibinfo
			{volume} {19}},\ \bibinfo {pages} {277} (\bibinfo {year} {1978})}\BibitemShut
	{NoStop}%
	\bibitem [{\citenamefont
		{Tikochinsky}(1979)}]{Hamilt_diagonalization_quadratik_Tikochinsky}%
	\BibitemOpen
	\bibfield  {author} {\bibinfo {author} {\bibfnamefont {Y.}~\bibnamefont
			{Tikochinsky}},\ }\href {\doibase 10.1063/1.524093} {\bibfield  {journal}
		{\bibinfo  {journal} {Journal of Mathematical Physics}\ }\textbf {\bibinfo
			{volume} {20}},\ \bibinfo {pages} {406} (\bibinfo {year} {1979})}\BibitemShut
	{NoStop}%
	\bibitem [{\citenamefont {Tzeng}\ \emph {et~al.}(1994)\citenamefont {Tzeng},
		\citenamefont {Ellis}, \citenamefont {Kuo},\ and\ \citenamefont
		{Osnes}}]{LMG-thermal_phase_partition_sum_Tzeng}%
	\BibitemOpen
	\bibfield  {author} {\bibinfo {author} {\bibfnamefont {S.~T.}\ \bibnamefont
			{Tzeng}}, \bibinfo {author} {\bibfnamefont {P.~J.}\ \bibnamefont {Ellis}},
		\bibinfo {author} {\bibfnamefont {T.}~\bibnamefont {Kuo}}, \ and\ \bibinfo
		{author} {\bibfnamefont {E.}~\bibnamefont {Osnes}},\ }\href@noop {}
	{\bibfield  {journal} {\bibinfo  {journal} {Nuclear Physics A}\ }\textbf
		{\bibinfo {volume} {580}},\ \bibinfo {pages} {277} (\bibinfo {year}
		{1994})}\BibitemShut {NoStop}%
	\bibitem [{\citenamefont {Hayn}\ and\ \citenamefont
		{Brandes}(2017)}]{hayn2017thermodynamics}%
	\BibitemOpen
	\bibfield  {author} {\bibinfo {author} {\bibfnamefont {M.}~\bibnamefont
			{Hayn}}\ and\ \bibinfo {author} {\bibfnamefont {T.}~\bibnamefont {Brandes}},\
	}\href@noop {} {\bibfield  {journal} {\bibinfo  {journal} {Physical Review
				E}\ }\textbf {\bibinfo {volume} {95}},\ \bibinfo {pages} {012153} (\bibinfo
		{year} {2017})}\BibitemShut {NoStop}%
	\bibitem [{\citenamefont {Wilms}\ \emph {et~al.}(2012)\citenamefont {Wilms},
		\citenamefont {Vidal}, \citenamefont {Verstraete},\ and\ \citenamefont
		{Dusuel}}]{wilms2012finite}%
	\BibitemOpen
	\bibfield  {author} {\bibinfo {author} {\bibfnamefont {J.}~\bibnamefont
			{Wilms}}, \bibinfo {author} {\bibfnamefont {J.}~\bibnamefont {Vidal}},
		\bibinfo {author} {\bibfnamefont {F.}~\bibnamefont {Verstraete}}, \ and\
		\bibinfo {author} {\bibfnamefont {S.}~\bibnamefont {Dusuel}},\ }\href@noop {}
	{\bibfield  {journal} {\bibinfo  {journal} {Journal of Statistical Mechanics:
				Theory and Experiment}\ }\textbf {\bibinfo {volume} {2012}},\ \bibinfo
		{pages} {P01023} (\bibinfo {year} {2012})}\BibitemShut {NoStop}%
	\bibitem [{\citenamefont {Dalla~Torre}\ \emph {et~al.}(2016)\citenamefont
		{Dalla~Torre}, \citenamefont {Shchadilova}, \citenamefont {Wilner},
		\citenamefont {Lukin},\ and\ \citenamefont {Demler}}]{dalla2016dicke}%
	\BibitemOpen
	\bibfield  {author} {\bibinfo {author} {\bibfnamefont {E.~G.}\ \bibnamefont
			{Dalla~Torre}}, \bibinfo {author} {\bibfnamefont {Y.}~\bibnamefont
			{Shchadilova}}, \bibinfo {author} {\bibfnamefont {E.~Y.}\ \bibnamefont
			{Wilner}}, \bibinfo {author} {\bibfnamefont {M.~D.}\ \bibnamefont {Lukin}}, \
		and\ \bibinfo {author} {\bibfnamefont {E.}~\bibnamefont {Demler}},\
	}\href@noop {} {\bibfield  {journal} {\bibinfo  {journal} {Physical Review
				A}\ }\textbf {\bibinfo {volume} {94}},\ \bibinfo {pages} {061802} (\bibinfo
		{year} {2016})}\BibitemShut {NoStop}%
	\bibitem [{\citenamefont {Brandes}(2008)}]{brandes2008waiting}%
	\BibitemOpen
	\bibfield  {author} {\bibinfo {author} {\bibfnamefont {T.}~\bibnamefont
			{Brandes}},\ }\href@noop {} {\bibfield  {journal} {\bibinfo  {journal}
			{Annalen der Physik}\ }\textbf {\bibinfo {volume} {17}},\ \bibinfo {pages}
		{477} (\bibinfo {year} {2008})}\BibitemShut {NoStop}%
	\bibitem [{\citenamefont {McCutcheon}\ \emph {et~al.}(2011)\citenamefont
		{McCutcheon}, \citenamefont {Dattani}, \citenamefont {Gauger}, \citenamefont
		{Lovett},\ and\ \citenamefont {Nazir}}]{mccutcheon2011a}%
	\BibitemOpen
	\bibfield  {author} {\bibinfo {author} {\bibfnamefont {D.~P.~S.}\
			\bibnamefont {McCutcheon}}, \bibinfo {author} {\bibfnamefont {N.~S.}\
			\bibnamefont {Dattani}}, \bibinfo {author} {\bibfnamefont {E.~M.}\
			\bibnamefont {Gauger}}, \bibinfo {author} {\bibfnamefont {B.~W.}\
			\bibnamefont {Lovett}}, \ and\ \bibinfo {author} {\bibfnamefont
			{A.}~\bibnamefont {Nazir}},\ }\href {\doibase 10.1103/PhysRevB.84.081305}
	{\bibfield  {journal} {\bibinfo  {journal} {Phys. Rev. B}\ }\textbf {\bibinfo
			{volume} {84}},\ \bibinfo {pages} {081305} (\bibinfo {year}
		{2011})}\BibitemShut {NoStop}%
\end{thebibliography}
\end{document}